\documentclass[fleqn,usenatbib]{mnras}
\usepackage{CJKutf8}
\usepackage{bm}
\usepackage{enumitem}
\usepackage[T1]{fontenc}
\usepackage{longtable}
\usepackage{tabu}
\usepackage{caption}
\usepackage{xcolor,colortbl}
\usepackage{pifont}
\usepackage{amsmath}
\DeclareRobustCommand{\VAN}[3]{#2}
\let\VANthebibliography\thebibliography
\def\thebibliography{\DeclareRobustCommand{\VAN}[3]{##3}\VANthebibliography}

\usepackage{listings}
\usepackage{xcolor}
\definecolor{codegreen}{rgb}{0,0.6,0}
\definecolor{codegray}{rgb}{0.5,0.5,0.5}
\definecolor{codepurple}{rgb}{0.58,0,0.82}
\definecolor{backcolour}{rgb}{0.95,0.95,0.96}

\lstdefinestyle{mystyle}{
    backgroundcolor=\color{backcolour},   
    commentstyle=\color{codegreen},
    keywordstyle=\color{magenta},
    numberstyle=\tiny\color{codegray},
    stringstyle=\color{codepurple},
    basicstyle=\ttfamily\footnotesize,
    breakatwhitespace=false,         
    breaklines=true,                 
    captionpos=b,                    
    keepspaces=true,                 
    numbers=left,                    
    numbersep=5pt,                  
    showspaces=false,                
    showstringspaces=false,
    showtabs=false,                  
    tabsize=2
}

\usepackage{graphicx}	
\usepackage{amsmath}	
\usepackage{amssymb}	
\usepackage{float}
\usepackage{newtxtext,newtxmath}
\usepackage{soul}
\usepackage{natbib}
\defcitealias{2023MNRAS.518.4249G}{Paper~I}
\usepackage[toc,page]{appendix}

\title[A catalogue of Galactic GEMS]{A catalogue of Galactic GEMS: Globular cluster Extra-tidal Mock Stars}

\author[S. M. Grondin et al.]{
Steffani M. Grondin,$^{1}$\thanks{E-mail: steffani.grondin@astro.utoronto.ca}
Jeremy J. Webb,$^{1,2}$
James M.M. Lane,$^{1}$ 
Joshua S. Speagle (\begin{CJK*}{UTF8}{gbsn}沈佳士\ignorespacesafterend\end{CJK*}),$^{3, 1, 4, 5}$
\newauthor 
\ Nathan W.C. Leigh$^{6, 7}$
\\
$^{1}$ David A. Dunlap Department of Astronomy \& Astrophysics, University of Toronto
50 St. George St, Toronto M5S 3H4, Canada \\
$^2$ Department of Science, Technology, and Society, Division of Natural Sciences, York University, 218 Bethune College, Toronto, ON, M3J1P3 \\
$^{3}$ Department of Statistical Sciences, University of Toronto, 9th Floor, Ontario Power Building, 700 University Ave, Toronto, ON M5S 3G3, Canada \\
$^{4}$ Dunlap Institute for Astronomy and Astrophysics, University of Toronto, 50 St George Street, Toronto, ON M5S 3H4, Canada \\
$^{5}$ Data Sciences Institute, University of Toronto, 17th Floor, Ontario Power Building, 700 University Ave, Toronto, ON M5G 1Z5, Canada \\
$^{6}$ Departamento de Astronom\'ia, Facultad Ciencias F\'isicas y Matem\'aticas, Universidad de Concepci\'on, Av. Esteban Iturra s/n Barrio Universitario,\\ Casilla 160-C, Concepci\'on, Chile \\
$^{7}$ Department of Astrophysics, American Museum of Natural History, New York, NY 10024, USA \\
}

\date{Accepted XXX. Received YYY; in original form ZZZ}

\pubyear{2024}

\begin{document}
\label{firstpage}
\pagerange{\pageref{firstpage}--\pageref{lastpage}}
\maketitle

\begin{abstract}
This work presents the Globular cluster Extra-tidal Mock Star (GEMS) catalogue of extra-tidal stars and binaries created via three-body dynamical encounters in globular cluster cores. Using the particle-spray code \texttt{Corespray}, we sample $N=50,000$ extra-tidal stars and escaped recoil binaries for 159 Galactic globular clusters. Sky positions, kinematics, stellar properties and escape information are provided for all simulated stars. Stellar orbits are integrated in seven different static and time-varying Milky Way gravitational potential models where the structure of the disc, perturbations from the Large Magellanic Cloud and the mass and sphericity of the Milky Way's dark matter halo are all investigated. We find that the action coordinates of the mock extra-tidal stars are largely Galactic model independent, where minor offsets and broadening of the distributions between models are likely due to interactions with substructure. Importantly, we also report the first evidence for stellar stream contamination by globular cluster core stars and binaries for clusters with pericentre radii larger than five kiloparsecs. Finally, we provide a quantitative tool that uses action coordinates to match field stars to host clusters with probabilities. Ultimately, combining data from the GEMS catalogue with information of observed stars will allow for association of extra-tidal field stars with any Galactic globular cluster; a requisite tool for understanding population-level dynamics and evolution of clusters in the Milky Way.
\end{abstract}



\section{Introduction}\label{sec:intro}

Globular clusters (GCs) are densely packed spherical collections of $10^{5}-10^{6}$ stars. Current observations find the existence of over $150$ GCs in our Milky Way, with complete orbital and structural parameters available in catalogues like \cite{1996AJ....112.1487H}, \cite{2018MNRAS.478.1520B} and \cite{Vasiliev2019}. Recent data releases from the European Space Agency satellite \textit{Gaia} \citep{2016A&A...595A...1G} and the Apache Point Observatory Galactic Evolution Experiment \citep[APOGEE;][]{2022ApJS..259...35A}, have also revolutionized our understanding of GC properties. For instance, \textit{Gaia} has allowed for exquisite and precise kinematic measurements of cluster stars \citep{2018A&A...616A..12G}, while APOGEE chemical abundances have been pivotal in identifying multiple stellar populations in GCs \citep[e.g.][]{2017MNRAS.466.1010S, 2019A&A...622A.191M, 2020MNRAS.492.1641M}. However, while much progress has been made to understand GCs in the present-day, probing cluster formation and evolution is challenging as GCs lose knowledge of their birth conditions as they lose stars and undergo structural changes over time.

\subsection{Three-body encounters in globular clusters}

Escaped stars from GCs inform a myriad of astrophysics; including cluster dynamics, binary fractions and star formation histories. GCs lose mass through a variety of mechanisms, with \cite{2023ApJ...946..104W} providing a comprehensive overview of stellar escape processes. While cluster evaporation and tidal stripping are the dominant channels for stellar escape from a GC, these methods primarily eject low-mass stars that have segregated to the outskirts of the cluster, yielding the creation of stellar streams or tidal tails. Since mass segregation causes high-mass stars and binaries to sink to a cluster's core, dynamical interactions in the core also must be considered to learn about high-mass stellar ejection, GC core evolution and the processes that ultimately produce isolated cluster field stars and binaries. 

One of the most common dynamical interactions in a GC core is a  \textit{three-body encounter}. Although the exact interaction rate depends on cluster core density and mass, three-body encounters typically occur once every $\sim 10$ Myr in a GC \citep{2011MNRAS.410.2370L}. For GC binary fractions greater than 10$\%$, three-body encounters will even occur more frequently than binary-binary interactions \citep{2011MNRAS.410.2370L}. In a three-body interaction, three stars will experience a close gravitational encounter, wherein a single star is ejected and the other two stars become a recoil binary system \citep[for a complete review of three-body dynamics, see][]{3body}. During the encounter, both the single star and recoil binary will experience a velocity kick as the binary hardens. The magnitude of the kick depends on the configuration of stellar masses, positions, and velocities, where stars can either be retained in the cluster or escape entirely \citep[e.g.][]{2023arXiv230913122L}. If the velocity kick is especially large and the stars travel beyond the tidal radius of the GC, the stars will become \textit{extra-tidal} in nature.

Since three-body interactions will continuously occur as the GC evolves, extra-tidal stars and binaries will be ejected at random times during a GC's orbit around the Galaxy. Coupled with the fact that the mass of the star/binary and the magnitude of the velocity kick will change for different three-body encounters, extra-tidal stars and binaries from a given cluster could be distributed all throughout the Milky Way. Unfortunately, the various initial configurations and interactions with Galactic sub-structures (e.g. orbiting through the Galactic disc or halo, experiencing dynamical perturbations from the Large Magellanic Cloud, etc.) make associating extra-tidal stars to birth clusters difficult. 

\subsection{Chemo-dynamical tagging to associate stars with clusters}

Fortunately, recent data releases by surveys like APOGEE have allowed for the advent of ``chemical tagging'': a method which utilizes chemical abundances to associate chemically similar stars with their birth environments. While chemical tagging has been an excellent tool to identify field stars that were likely born in the same birth cluster \cite[e.g.][]{2018MNRAS.473.4612K, 2020MNRAS.496.5101P}, chemistry alone does not definitively indicate that the stars originated in the same GC \citep{2018ApJ...853..198N, 2019ApJ...883..177N, 2021A&A...654A.151C}. 
However, since most stars in a GC are born from the same giant molecular cloud, they will not just share similar abundances, but kinematics as well. Thus, ``\textit{chemo-dynamical} tagging'' (associating stars via their chemistry \textit{and} stellar kinematics) is a useful tool to identify stellar birth environments with higher confidence \citep[e.g.][]{2018ApJ...860...70C, 2020A&A...637A..98H, 2020MNRAS.494.2268W}.

Chemo-dynamical tagging has also been used to identify extra-tidal field stars of GCs, with \cite{2020ApJ...900..146C} combining APOGEE DR14 abundances and radial velocities to identify new extra-tidal stars around the Galactic GCs M53 and NGC 5053. In \cite{2023MNRAS.518.4249G}, we developed a new method to identify extra-tidal stars produced via three-body encounters in GC cores, using M3 as a case study. With APOGEE DR17 chemistry and radial velocities, \cite{2023MNRAS.518.4249G} identified 103 extra-tidal candidates of M3 through application of the t-Stochastic Neighbour Embedding \citep[t-SNE;][]{JMLR:v9:vandermaaten08a} and Uniform Manifold Approximation and Projection \citep[UMAP;][]{2018arXiv180203426M} dimensionality reduction algorithms. To confirm whether each candidate indeed originated from a three-body encounter in M3's core, we developed \texttt{Corespray}: a Python-based particle spray code simulating extra-tidal stars from Galactic GCs. \texttt{Corespray}'s ability to quickly sample three-body interactions makes it an ideal tool to investigate the spatial, kinematic and stellar characteristics of dynamically-created extra-tidal stars and binaries on a large scale. 

While \cite{2023MNRAS.518.4249G} were successful at identifying 10 new extra-tidal stars of one Galactic GC, a catalogue containing mock extra-tidal star and binary data for \textit{all} Galactic GCs allows for an even deeper exploration into the evolution of Galactic GCs and the Milky Way as a whole. For instance, kinematic distributions could be used to associate field extra-tidal stars with birth clusters, allowing for identification of extra-tidal single star and binary pairs. Moreover, \cite{2023ApJ...953...19C} suggest that dynamical interactions in GC cores could yield hyper-velocity or runaway stars in the Milky Way, where a catalogue of extra-tidal stars could be used to trace observed stars back to clusters.

\subsection{This work}

In this study, we present the Globular cluster Extra-tidal Mock Star (GEMS) catalogue. The GEMS sample represents the first simulated mock catalogue of extra-tidal stars and binaries created via three-body encounters in Galactic GC cores. We simulate extra-tidal star and binary spatial (on-sky positions, distances), kinematic (proper motions, radial velocities, action coordinates) and stellar information (masses, escape velocities) for all 159 Galactic GCs listed in \cite{2018MNRAS.478.1520B} using \texttt{Corespray}. In Section \ref{sec:methods}, we describe \texttt{Corespray} and outline the initial simulation set-up and input parameters considered in this study. Spatial and kinematic distributions for both the single extra-tidal stars and recoil binaries of all GCs are presented in Section \ref{sec:results}, with the results discussed in Section \ref{sec:discussion}. Section \ref{sec:discussion} also presents a new quantitative tool that uses extra-tidal star action coordinates to match field stars with host GCs, where the code is available in Appendix \ref{sec:association}. In combination with chemo-dynamical data from APOGEE and \textit{Gaia}, the GEMS catalogue aims to act as a resource for understanding population-level dynamics and evolution of GCs in the Milky Way.

\section{Simulating extra-tidal stars with 
\texttt{Corespray}}\label{sec:methods}

\subsection{The \texttt{Corespray} particle spray code}

To construct our extra-tidal star and binary distributions, we use the particle spray code \texttt{Corespray}\footnote{For a complete outline of how \texttt{Corespray} works, please refer to \cite{2023MNRAS.518.4249G} or visit \href{https://github.com/webbjj/corespray}{https://github.com/webbjj/corespray} for download and usage instructions.} \citep{2023MNRAS.518.4249G}. Using the theoretical three-body dynamics framework presented in \cite{3body}, \texttt{Corespray} samples three-body interactions in GC cores until $N$ single stars have travelled beyond the cluster's tidal radius. Running \texttt{Corespray} requires the definition of GC orbital and structural parameters. Specifically, each cluster's orbit is defined from its sky position, distance, proper motion and radial velocity. Structural parameters (i.e. mass, tidal radius, core density, velocity dispersion and escape velocity) are also essential for initializing each cluster and its unique escape criteria. Importantly, \texttt{Corespray} also takes in parameters like cluster orbital period, a number of extra-tidal stars to sample, a separation radius for core stars that experience three-body encounters, a stellar mass function and a model for the Galactic potential. Each of these parameters are defined by the user. 

Once the aforementioned initial conditions are inputted, \texttt{Corespray} generates extra-tidal stars at random times along a cluster's orbit throughout the Galaxy. Spatial (e.g. right ascension and declination, distance), kinematic (e.g. proper motions, radial velocities, action coordinates) and stellar (e.g. masses) information is produced for each single and binary escaper, allowing orbits to be integrated with \texttt{galpy} \citep{2015ApJS..216...29B} and large distributions of escapers to be produced.

While N-body integrators like \texttt{Nbody6} \citep{Aarseth2003}, \texttt{$\phi$-GRAPE} \citep{Harfst2007}, and \texttt{PeTar} \citep{Wang2020} directly model three-body interactions in cluster cores, simulating the evolution of realistic globular clusters is computationally expensive due to their large particle numbers and old ages \citep{Wang2016}. Complicated by the fact that the birth properties of a given GC are needed to simulate its evolution to the present-day \citep{Heggie2014}, the prospect of producing a suite of direct $N$-body simulations of the Milky Way's GC population remains daunting. To study the outcomes of three-body interactions within GCs, many randomizations of each Galactic cluster would need to be completed, adding further to the study's computational expense. \texttt{Corespray}’s abilities to quickly sample and integrate orbits for extra-tidal stars and binaries of any Galactic GC makes it an ideal tool to generate statistically significant distributions of escaped stars throughout the Milky Way.

\subsection{Generating mock extra-tidal stars} \label{sec:corespray}

\subsubsection{Defining globular cluster parameters} \label{sec:gcsetup}

In this study, GC structural parameters are obtained from the \cite{2018MNRAS.478.1520B} GC catalogue \footnote{All fundamental Galactic GC parameters can be accessed via the online database at \href{https://people.smp.uq.edu.au/HolgerBaumgardt/globular/}{https://people.smp.uq.edu.au/HolgerBaumgardt/globular/}.}. All GC orbital parameters are obtained from \cite{2019MNRAS.482.5138B}, \cite{2021MNRAS.505.5957B} and \cite{2021MNRAS.505.5978V}. To define a tidal radius ($r_{t}$) boundary (and thus a threshold for where the star or binary will have officially "escaped" the cluster), we use the \cite{2013ApJ...764..124W} $r_{t}$ at perigalacticon to compute the $r_{t}$ at apogalacticon. This definition ensures that we are only considering stars that are beyond $r_{t}$ at \textit{all} points in the GC's orbit through the Galaxy as being extra-tidal. Furthermore, since \texttt{Corespray} uses a combined Galactic and King potential to encompass influences from both the Milky Way and the GC itself, we must define a King-model central potential parameter ($W0$) for each GC. From \cite{1996AJ....112.1487H}, the King-model central concentration parameter ($c$) can be obtained using the previously defined $r_{t}$ and core radii ($r_{c}$) presented in \cite{2018MNRAS.478.1520B} via $c=\log_{10}(r_{t}/r_{c})$. The conversion from $c$ to $W0$ is then performed using \texttt{clustertools}\footnote{For a complete description of \texttt{clustertools} capabilities and installation instructions, please visit \href{https://github.com/webbjj/clustertools}{https://github.com/webbjj/clustertools}.} \citep{webbjj_2022_6656647} and data from \texttt{gridfit} \citep{McLaughlin2008}.

Another essential GC parameter to consider is the time period over which extra-tidal stars are simulated. Contrary to \cite{2023MNRAS.518.4249G} who searched for \textit{recently} escaped extra-tidal stars of M3, this study wishes to examine if/how the properties of extra-tidal stars and binaries evolve as a \textit{function of time}. As such, for the purposes of this catalogue we generate stars that escape during random times along five orbital periods (P\textsubscript{orb}) of the GC through the Galaxy. The choice to simulate 5 orbital periods of escape is motivated by computational expense, as simulation time must be balanced against the number of extra-tidal stars being generated \footnote{Simulating escape over more than 5 orbital periods is entirely possible within \texttt{Corespray} and should be considered for users interested in longer escape times.}. To determine the P\textsubscript{orb} of each GC, we compute both the radial and azimuthal orbital periods using \texttt{galpy} \citep{2015ApJS..216...29B}. Each P\textsubscript{orb} used in this study represents the maximum of the two computed cluster periods. It is important to note that 14 GCs in our sample (AM 1, Arp 2, BH 140, Crater, Eridanus, NGC 2419, Pal 3, Pal 12, Pal 14, Pyxis, Sagittarius II, Ter 7, Ter 8 and Whiting 1) have P\textsubscript{orb} $> 2$ Gyr. Consequently, this could result in stellar escape times where t\textsubscript{esc} $= 5 \times$ P\textsubscript{orb} corresponds to times older than the age of the Universe. To ensure our distributions for these clusters are physical, we impose upper escape time boundaries of t\textsubscript{esc} $< 12$ Gyr, where 12 Gyr represents an intermediate age for the GCs in the Milky Way \citep{Forbes2010}.

It is important to note that in this study, we assume a cluster's properties remain constant over the course of five orbital periods. However for GCs with long orbital periods, this assumption is incorrect. Cluster cores get denser as they evolve towards core collapse \citep{2003gmbp.book.....H}, so it is likely that clusters had lower central densities in the past. Lower central densities and velocity dispersions would result in three-body interactions that yield weaker kicks, potentially causing our distributions of extra-tidal stars to be wider for stars that escape later. This situation is of course complicated by the evolution of a cluster's binary star population and tidal field. Hence, for an in-depth analysis of a specific GC's extra-tidal star history, one may consider pairing \texttt{Corespray} with a cluster's evolution track as generated via a fast-code prescription \citep[e.g. the \texttt{Evolve Me A Cluster of StarS (EMACSS)} code;][]{2012MNRAS.422.3415A} or a completed simulation of the internal dynamics \citep[e.g. the \texttt{Cluster Monte Carlo (CMC)} code;][]{2022ApJS..258...22R}.

\subsubsection{Defining three-body encounter parameters}

In addition to GC structural and orbital parameters, \texttt{Corespray} offers a variety of other tunable inputs to address specific science questions. We outline our choices for (i) the number of extra-tidal stars simulated, (ii) the binding energy of the binary (iii) the separation radius between the binary and single star during the three-body encounter and (iv) the stellar initial mass function (IMF) below.

First, since a goal of this study is to produce complete parameter spaces of extra-tidal stars and their recoil binaries, we choose to sample $N=50,000$ single escapers for each GC. Importantly, the fraction of recoil binaries that escape a cluster is dependent on individual GC characteristics and masses of the stars in the three-body encounter. Thus, high-$N$ distributions allow us to completely explore the parameter spaces of escaped single extra-tidal stars, while also producing a statistically significant sample of escaped recoil binaries. It is important to note here that in generating a statistically significant sample of ET stars we do not assume or calculate a three-body interaction rate for the GCs. Sub-sampling the distribution of $N=50,000$ ET stars based on a given clusters three-body interaction rate would allow for a single realization of where a given clusters ET stars could be located.

Second, in order for the three-body interactions simulated with \texttt{Corespray} to reflect the interactions that occur in GCs, we assume a cluster's binary star binding energy distribution follows Opik's Law (a distribution with power-law slope of -1). Hence, interactions with soft binaries will be sampled more frequently than interactions with hard binaries, which reflects what is expected in actual GCs \citep{Leigh2022}.

Third, to allow for different configurations of three-body encounters, we sample a variety of separation radii between the single star and binary during each interaction. Specifically, we randomly sample separation radii between the semi-major axis of the binary and twice the mean separation of stars in the GC's core. Since only encounters that lead to escaped stars are modelled, setting such a large upper bound on the separation radius ensures we cover the entire range of possible interactions albeit with increased computational expense.

Finally, initializing three-body encounters requires assigning masses to each component of the system. While sampling from a IMF allows us to probe a variety of mass configurations, defining an IMF to model mass distributions in GC cores is challenging. This is because (i) the high stellar core number density makes it difficult to observe individual stars and (ii) many clusters contain dark compact objects (i.e. neutron stars and black holes), which despite significantly contributing to the mass distribution, would remain observationally unresolved. Furthermore, mass segregation and tidal stripping over billions of years would remove many low mass cluster stars, shifting the core mass distribution to an overall higher average stellar mass. To mitigate some of these challenges, we sample masses from a  \cite{2001MNRAS.322..231K} IMF containing $N=10^{6}$ stars. This IMF is then evolved for 12 Gyr with \texttt{McLuster} \citep{Kupper2011} to incorporate some of the aforementioned effects and ultimately obtain a realistic estimate for GC stellar mass distributions at the present-day. It is important to note that the components of the binaries will also evolve towards a top-heavy IMF over time, which is not accounted for in this set-up. While this is a caveat of the initialized three-body encounters, the GEMS catalogue could be sampled to focus on stars over a given mass range, allowing for an examination into the results of interactions featuring heavier binary mass configurations. 

\subsection{Galactic potential models} \label{sec:models}

In the last few decades, significant attention has been paid towards understanding our Milky Way's gravitational potential \citep[e.g.][]{2010ApJ...712..260K, Irrgang2013, Bovy2013, 2015ApJS..216...29B, 2017MNRAS.465...76M, 2019MNRAS.486.2995M, Eadie2019}. The precise potential of the Galaxy still remains an open question \citep[for a comprehensive discussion of the structural properties of the Milky Way, see][]{2016ARA&A..54..529B}, where different components of the Galaxy can influence the orbits of stars in a variety of different ways. For instance, recent observations have shown that non-axisymmetric structures like a bar can play an important role of the dynamics and structure of clusters and stars in and around the Galactic disc \citep[e.g.][]{2016MNRAS.460..497H, 2016ApJ...824..104P, 2017NatAs...1..633P, 2019MNRAS.484.2009B, 2023arXiv230905733T}. Furthermore, studies have shown that the Large Magellanic Cloud could cause perturbations to the Milk Way halo \citep[e.g.][]{2019ApJ...884...51G, 2021MNRAS.506.2677E}, additionally impacting the orbits of stars. Moreover, the mass and shape of the halo -- where GCs primarily spend their time -- are poorly constrained. Hence, it is essential to consider a variety of potential models when exploring the locations and kinematics of GCs and their extra-tidal stars over long periods of time. In this study, we simulate extra-tidal stars/binaries in \textit{seven different potential models}:

\begin{enumerate}[leftmargin=.5em]
    \item A 3-component static potential \citep[\texttt{MWPotential2014} in][] {2015ApJS..216...29B} 
     \vspace{-.2cm}
    \item \texttt{MWPotential2014} + a disc with a rotating bar and transient-wave spiral arms
    \vspace{.2cm}
    \item A 6-component static potential with a heavy halo \citep[\texttt{McMillan17} in][]{2017MNRAS.465...76M}
    \vspace{.2cm}
    \item \texttt{MWPotential2014} + varying halo shapes:
        \begin{itemize}[]
        \vspace{-.2cm}
        \setlength{\itemindent}{1em}
            \item Oblate halo, $q=0.9$
            \item Spherical halo, $q=1.06$
            \item Prolate halo, $q=1.30$
        \end{itemize}
    \vspace{-.05cm}
    \item \texttt{MWPotential2014} + infall of the Large Magellanic Cloud
\end{enumerate}

Each Milky Way model is described in detail below, however for complete model descriptions, please refer to the referenced original studies and codes. 

\subsubsection{Baseline: a static tidal field} \label{sec:pot-baseline}

A simple model for the Milky Way's potential is \texttt{MWPotential2014} \citep{2015ApJS..216...29B}. \texttt{MWPotential2014} contains three main components: an exponentially cut-off power-law density profile to model the Galactic bulge, a \cite{1975PASJ...27..533M} disc and a Navarro-Frenk-White dark matter halo \citep{1996ApJ...462..563N}, with the model itself fit to Milky Way data \citep{2015ApJS..216...29B}. Despite its utility, the default version of this model (i) is time-independent, (ii) contains a lower than expected halo mass based on recent observations \citep{2019A&A...621A..56P, 2021MNRAS.501.5964D, 2022MNRAS.516..731B} and (iii)  has a slightly lower circular velocity at the Solar radius than preferred by modern measurements \citep[e.g.][]{2019ApJ...871..120E}. Since we are interested in examining extra-tidal star characteristics across a variety of orbital periods and locations in the Galaxy, additional investigation into large scale structure, time-dependent potential evolution and halo structure is necessary. Regardless,  \texttt{MWPotential2014} is used as a baseline in this study. 

\subsubsection{Disc: a rotating bar + transient-wave spiral arms} \label{sec:pot-disc-transient}

Although GCs spend much of their time in the Galactic halo, interactions with the Galactic disc also occurs during their orbits through the Galaxy. As extra-tidal stars and binaries are ejected at random times during a cluster's orbit, it is possible that large-scale interactions with the disc could influence an extra-tidal star's orbital properties. While direct observational evidence for a central bar at the centre of the Milky Way has been known for many decades \citep{1991ApJ...379..631B}, the formation and evolution of the Galaxy's spiral arms is more uncertain. Historically, it has been thought that the spiral arms of the Milky Way originally formed through ``density waves'' -- overdense, rotating regions of space \citep[for an in-depth review of the history of spiral structure, see][]{2016ARA&A..54..667S}. However, recent observational studies have favoured \textit{transient-wave} spiral arms, which are spiral arms that grow and decay \textit{over time} with radially dependent rotational rates. For instance,  \cite{2021A&A...652A.162C} use \textit{Gaia} EDR3 data to find that different pattern speeds are better aligned with transient spirals than typical density waves. 

As such, we modify our baseline \texttt{MWPotential2014} potential to incorporate a rotating Galactic bar from \cite{2000AJ....119..800D}, with evolving spiral arms from \cite{2019MNRAS.490.1026H} -- henceforth known as the ``Bar+Transient Wave'' potential\footnote{This Bar+Transient Wave potential model featuring large-scale structure in the Milky Way disc has been integrated into \texttt{Corespray}, where further description of how to implement this time-dependent potential is available at \href{https://github.com/webbjj/corespray/tree/main/corespray/potential}{https://github.com/webbjj/corespray/tree/main/corespray/potential.}}. Notably, this potential model is \textit{time-dependent}, where the first two transient arms are initialized at a time equal to 1 Gyr before the start of the simulation. One arm is initialized with an amplitude of zero in density-space while the other arm is at maximum amplitude. The spiral arms then continue to grow and decay over the evolution of the cluster. The lifetime of a given arm is 460 Myr and the initialization is such that there are always two active arms, with one reaching maximum every 230 Myr \citep{2019MNRAS.490.1026H}.

\begin{figure*}
    \centering
    \includegraphics[width=0.83\textwidth]{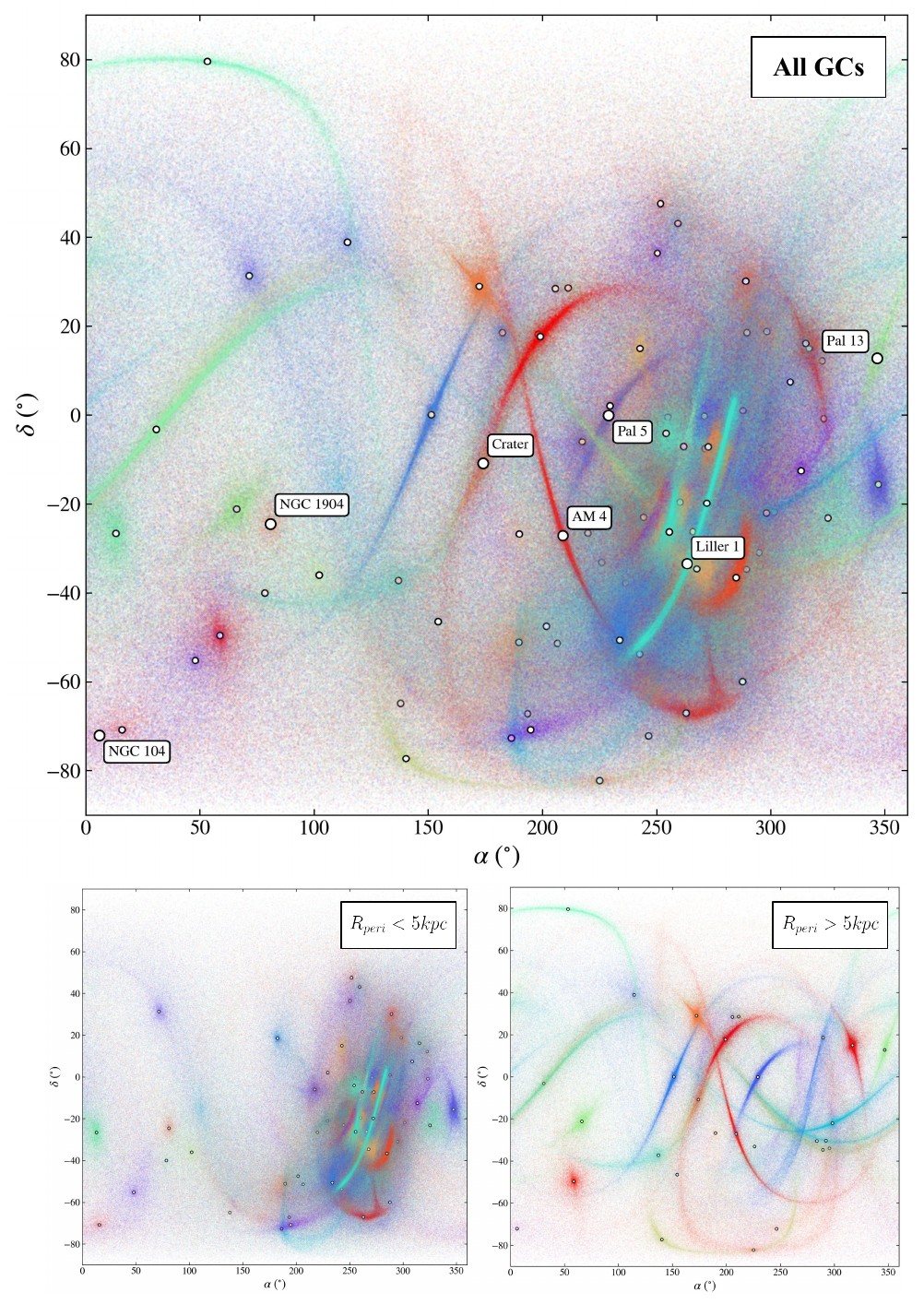}
    \caption{Spatial distributions (right ascension $\alpha$ and declination $\delta$) of single extra-tidal stars from three-body encounters in GC cores. $50,000$ extra-tidal stars are simulated for each cluster with \texttt{Corespray} \citep{2023MNRAS.518.4249G}, with orbits integrated in our baseline Galactic potential model \texttt{MWPotential2014} \citep{2015ApJS..216...29B} described in Section \ref{sec:pot-baseline}. Coloured points represent individual GC extra-tidal stars, where the white circles show the centre of each cluster. GCs specifically discussed in this study are labelled, with their cluster centres highlighted by large white circles. The top panel shows the spatial distributions of extra-tidal stars for all 159 GCs in \citet{2018MNRAS.478.1520B}, whereas the bottom left and right panels show the distributions for stars of GCs with pericentre radii $ R_{\text{peri}}<5$ kiloparsecs and $R_{\text{peri}}>5$ kiloparsecs, respectively.}
    \label{fig:spatial}
\end{figure*}

\subsubsection{Halo: a static tidal field with a heavy dark matter halo} \label{sec:pot-mcmillan}

As many GCs spend the majority of their orbits in the Galactic halo, it is important to consider a range of halo potentials when simulating extra-tidal stars and binaries. Unfortunately, constraining the mass of the Galactic dark matter halo is difficult, as measurements are typically extrapolations from the luminous matter in our Galaxy. To incorporate a more complex gravitational potential that features a more realistic halo mass, we use the \cite{2017MNRAS.465...76M} model. This potential models the Milky Way using six components (as opposed to three in \texttt{MWPotential2014}) and is fit to modern kinematic data. Importantly, the \cite{2017MNRAS.465...76M} potential model contains a larger virial mass compared to \texttt{MWPotential2014} ($1.3 \times 10^{12} M_{\odot}$ versus $0.8 \times 10^{12} M_{\odot}$, respectively). Conveniently, the \cite{2017MNRAS.465...76M} potential is also implemented in \texttt{galpy} \citep{2015ApJS..216...29B} as \texttt{McMillan17}.

\subsubsection{Halo: a non-spherical dark matter halo} \label{sec:nonspherical}

We also examine whether our extra-tidal star distributions are influenced by the triaxiality of the Milky Way’s dark matter halo. The shape of the dark matter halo is typically quantified by the flattening parameter $q$, the ratio of the axis scale length in the disk plane versus the polar axis ($c/a$). A value of $q=1$ represents complete sphericity of the dark matter halo, whereas $q<1$ implies an oblate halo and $q>1$ corresponds to a prolate halo. The Milky Way's dark matter halo shape is currently poorly constrained, thus we consider three different values of $q$ that span a variety of $q$ values: (i) $q=0.9$ \citep[][who determine $q$ by fitting properties of the Sagittarius stellar stream]{2013ApJ...773L...4V}, (ii) $q=1.06$ \citep[][who determine $q$ via the stellar streams of NGC3201, M68, and Palomar 5]{2023MNRAS.tmp.1860P} and (iii) $q=1.30$ \citep[][who determine $q$ via proper motions of 75 GCs from \textit{Gaia} DR2 and 25 GCs from the Hubble Space Telescope]{2019A&A...621A..56P}. For simplicity, we do not consider disk-plane non-axisymmetry ($b/a \neq 1$) or tilting of the dark matter halo. This assumption is reinforced by theoretical arguments which posit that baryonic disk growth encourages the dark halo to align with the disk plane and become axisymmetric, a picture backed up by evidence from simulations \citep[e.g.][]{dubinski94,abadi10,kazantzidis10}.

Once again, we define a 3-component Galactic potential model, where the bulge and disc components are the same as in our baseline (\texttt{MWPotential2014}), however we adopt a \texttt{TriaxialNFWPotential} (\texttt{normalize=0.35}) with default components in \texttt{galpy} \citep{2015ApJS..216...29B}. The only thing that is varied between each run is the flattening parameter $q$. Since each $q$ value corresponds to a new Galactic potential model, a subset of three GCs is chosen to probe a representative range of Galactocentric distances while minimizing computational expense. Since triaxiality effects would be stronger for clusters located in the Galactic halo, we select \citep[from][]{2018MNRAS.478.1520B} halo clusters M3 ($d\sim10$ kpc), Palomar 5 ($d\sim22$ kpc) and Pyxis ($d\sim36$ kpc) as our test GCs.

\subsubsection{Halo: perturbations from the Large Magellanic Cloud} \label{sec:pot-lmc}

Recent observations and simulations have shown that the Large Magellanic Cloud (LMC) exhibits a gravitational pull on the Milky Way, causing both a dynamical wake and non-inertial Galactocentric reference frame \citep[e.g.][]{2019ApJ...884...51G, 2021ApJ...919..109G, 2021Natur.592..534C, 2021MNRAS.506.2677E, galaxies11020059}. Moreover, these studies suggest that the infall of the LMC could cause dynamical perturbations to the Milky Way's stellar and dark matter halos, causing GCs -- and ejected stars thereof -- orbiting near the LMC to experience kinematic perturbations. Thus, we define a non-inertial frame force due to the LMC in \texttt{galpy} \citep{2015ApJS..216...29B}, where the exact model used is described in detail at \href{https://docs.galpy.org/en/v1.7.2/orbit.html#orbit-example-lmc-dynfric}{https://docs.galpy.org/en/v1.7.2/orbit.html\#orbit-example-lmc-dynfric} and added to our baseline \texttt{MWPotential2014}.

\section{Results}\label{sec:results}

As outlined in Section \ref{sec:methods}, we simulate 50,000 extra-tidal stars and corresponding recoil binaries in five different Galactic potential models for 159 GCs in \cite{2018MNRAS.478.1520B} using \texttt{Corespray}. Each extra-tidal star and binary contains complete spatial (e.g. right ascension, declination, distance), kinematic (e.g. proper motions and radial velocities) and stellar (e.g. masses) information, allowing for an investigation into an array of different characteristics. Escape times and velocities of each star are also known, allowing for an examination into how the extra-tidal distributions change with time. Below, we highlight the outcomes of extra-tidal stars/binaries simulated in the baseline static tidal field (\texttt{MWPotential2014}), where the catalogue itself is available online at \href{https://zenodo.org/record/8436703}{https://zenodo.org/record/8436703}. Note that we ignore observational uncertainties on the cluster phase-space coordinates, however one could account for this by running several iterations of a specific cluster with \texttt{Corespray}.

\subsection{Extra-tidal single star properties}

\subsubsection{Spatial distributions} \label{sec:spatial}

To examine how extra-tidal stars from three-body encounters in GC cores are distributed throughout the Milky Way, we show the \texttt{Corespray} spatial distributions of all single extra-tidal stars for each cluster in \cite{2018MNRAS.478.1520B} in Figure \ref{fig:spatial}. It is clear that extra-tidal stars can end up far from their original parent cluster, with distribution shapes widely varying. Elongated extra-tidal star streams are typically observed in clusters that have pericentre radii ($R_{\text{peri}}$) larger than five kiloparsecs (bottom right panel of Figure \ref{fig:spatial}). Of the 31 clusters in \cite{2018MNRAS.478.1520B} that have $R_{\text{peri}} > 5$kpc, more than half have extremely low cluster escape velocities, with $v_{esc} < 5$km/s. Generally, these clusters also have lower core densities and higher core radii compared to clusters with $R_{\text{peri}} < 5$kpc, potentially allowing the extra-tidal stars to more easily escape the cluster altogether. This stream-like behavior is contrary to clusters with $R_{\text{peri}} < 5$kpc (bottom left panel of Figure \ref{fig:spatial}), where extra-tidal stars typically congregate around each GC's centre. Clusters with small $R_{\text{peri}}$ typically have shorter orbital periods compared to clusters in the halo. Hence, stars that escape these clusters will drift away from the progenitor cluster faster and are more likely to end up dispersed all throughout the Galaxy. Thus, the only structure that is observed for these clusters is a concentration around the cluster itself. In combination with GC structural properties, the cluster's orbit and subsequent interaction with Galactic substructure could also influence the overall shape of each extra-tidal star distribution.

\subsubsection{Action distributions} \label{sec:actions}

\begin{figure*}
    \centering
    \includegraphics[width=0.95\textwidth]{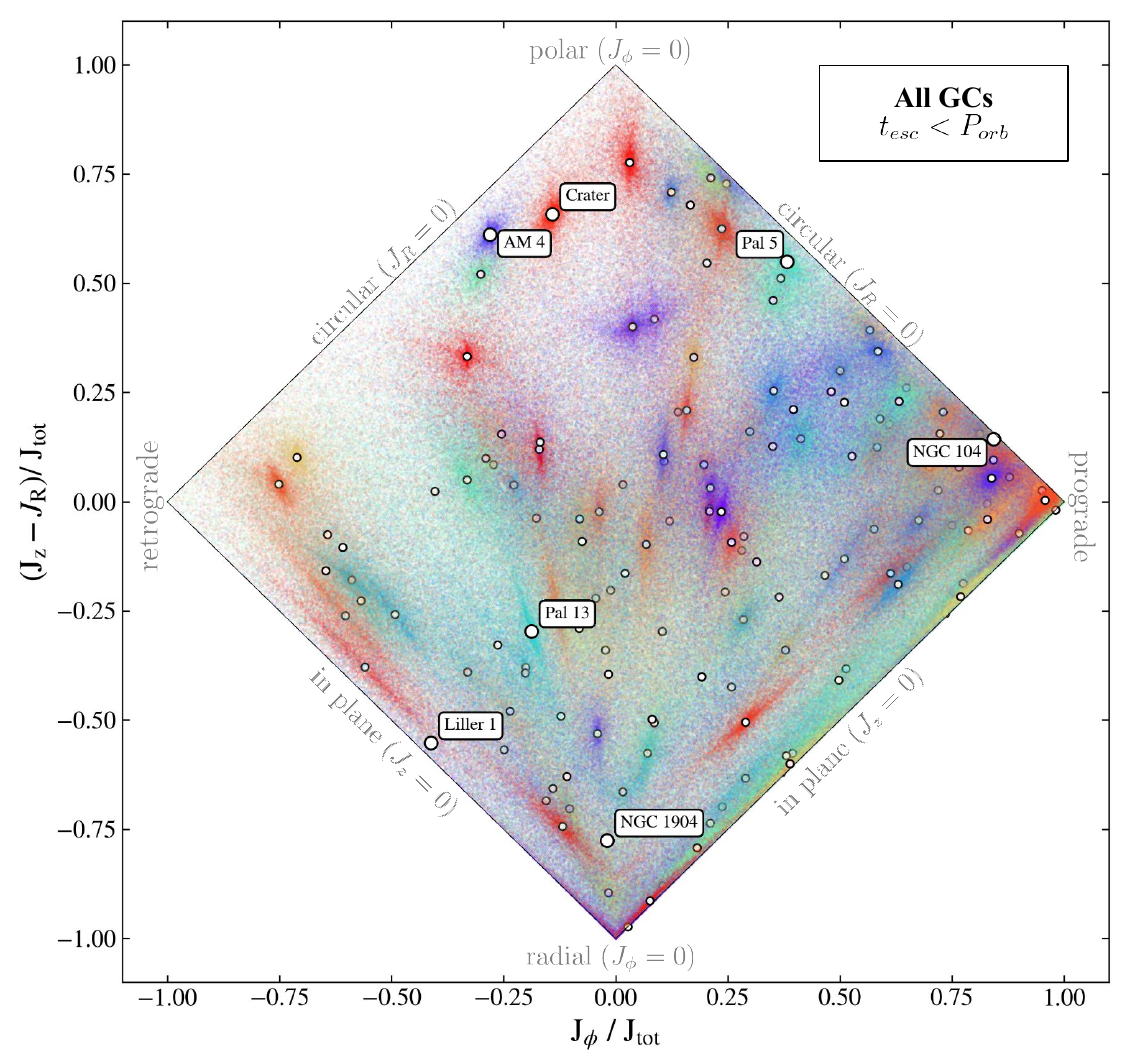}
    \caption{The action diamond space for single extra-tidal stars of all 159 GCs in \citet{2018MNRAS.478.1520B}. Coloured points represent individual GC extra-tidal stars, where the white circles show the centre of each cluster. GCs specifically discussed in this study are labelled, with their cluster centres highlighted by large white circles. For clarity, we only plot the actions of stars that escaped the clusters up to one orbital period in the past, as the distributions generally tend to broaden when all 50,000 stars are included. All actions are computed using 6-D information ($\alpha$, $\delta$, $d$, $\mu_{\alpha}$, $\mu_{\delta}$, $v_{los}$) from our \texttt{Corespray} simulations in a baseline \texttt{MWPotential2014} potential model with the St\"{a}ckel approximation in \texttt{galpy} \citep{2015ApJS..216...29B} (see Appendix \ref{sec:Staeckel} for a discussion of how calculated actions can depend on orbital phase in static potentials when using the St\"{a}ckel approximation). Annotations of the diamond are adapted from Figure 5 in \citet{2019MNRAS.488.1235M} and Figure 5 in \citet{Vasiliev2019}.}
    \label{fig:actions}
\end{figure*}

\begin{figure*}
    \centering
    \includegraphics[width=0.95\textwidth]{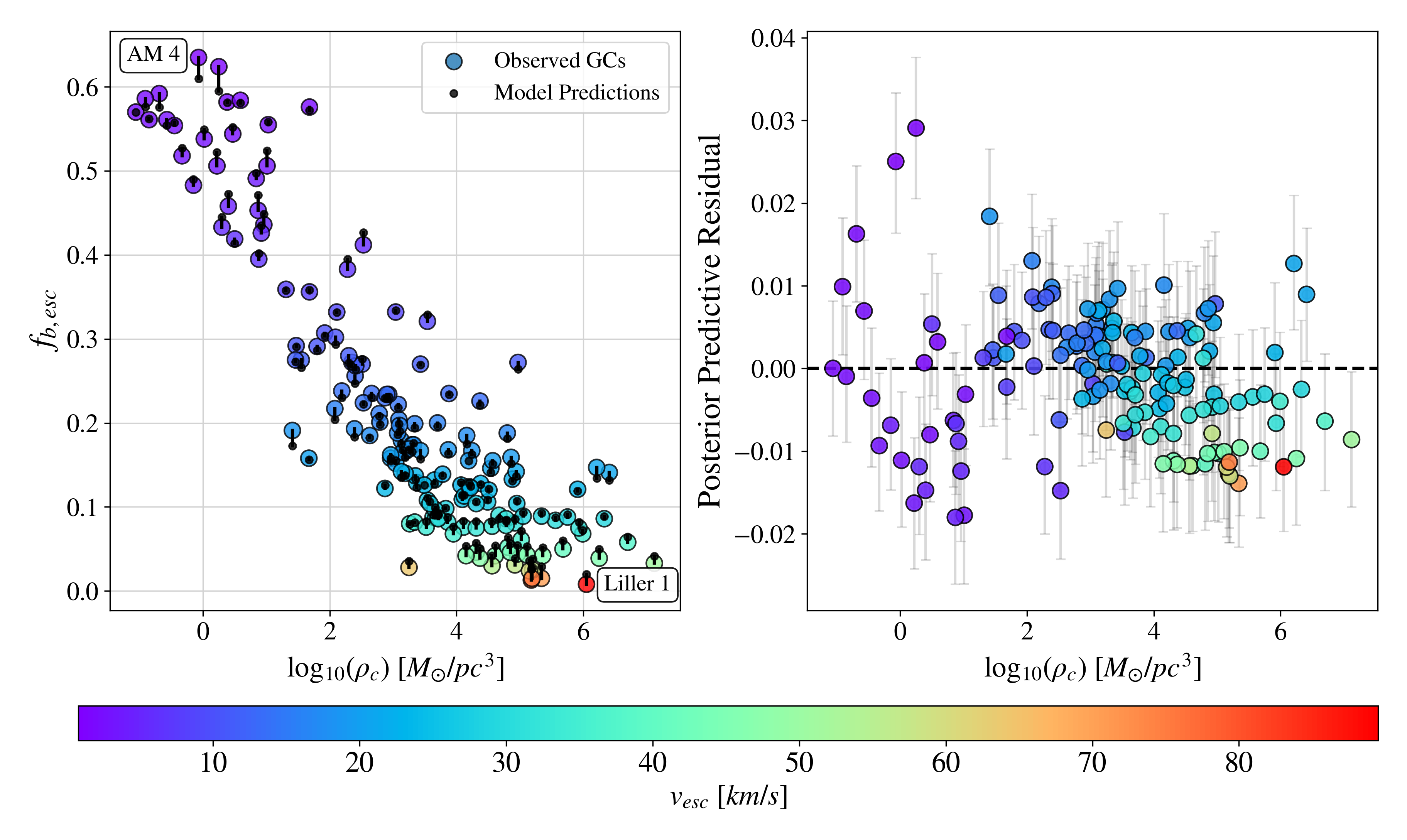}
    \caption{The recoil binary escape fraction ($f_{b, esc}$) as a function of cluster core density ($\rho_{c}$) and escape velocity ($v_{esc}$) for all 159 GCs in \citet{2018MNRAS.478.1520B}. \textit{Left panel:} GCs in \citet{2018MNRAS.478.1520B} are highlighted as coloured circles. We model $f_{b, esc}$ as a modified logistic function that depends on the (logarithm of) $\rho_{c}$ and $v_{esc}$ and determine the model $f_{b, esc}$ values for each cluster (black points) using the posterior predictives computed from \texttt{dynesty} nested sampler \citep{2020MNRAS.493.3132S}. The cluster with the largest binary escape fraction is AM 4 ($f_{b, esc} = 63.5\%$) and the cluster with the smallest binary escape fraction is Liller 1 ($f_{b, esc} = 0.8\%$). Generally, the number of binaries that escape the cluster significantly increases as $\rho_{c}$ increases and $v_{esc}$ decreases. \textit{Right panel:} The residuals between the \texttt{Corespray} $f_{b, esc}$ values and the model posterior predictive $f_{b, esc}$ values (coloured points). Error bars are computed using the median and the 16th/84th percentiles derived from the \texttt{dynesty} results.} 
    \label{fig:binary-frac}
\end{figure*}

Generally, GC stars have similar kinematic properties (i.e. proper motions, radial velocities), as they are born from the same collapsing and rotating giant molecular cloud. However extra-tidal stars are the result of dynamical interactions, which impart velocity kicks onto each escaper and thus change the overall kinematics of the systems. Post-cluster escape, extra-tidal stars can also interact with large scale Galactic substructure, further altering the kinematics of the stars. Combined with unique three-body encounter configurations (i.e. masses and initial separations) and locations of ejection, extra-tidal stars from one GC could thus exhibit a range of kinematics, often much different than those of the host cluster itself. To further complicate things, proper motions and radial velocities of \textit{both} a GC and its escaped stars will change with orbital phase, making it difficult to establish star-cluster associations using \textit{non-conserved} kinematic quantities alone.

To navigate this issue, we compute actions -- canonical momenta which are, in principle, conserved along orbit trajectories \citep[see Chapter 3.5 of][ for a review]{2008gady.book.....B} -- for our samples of extra-tidal stars. Actions $J_{R}$ and $J_{z}$ (Note $J_{\phi}=L_\mathrm{z}$ is trivial) are computed using the ``St\"{a}ckel fudge'' technique of \citet{Binney2012}, implemented in \texttt{galpy} as described by \citet{Mackereth2018}. For each star, we compute the appropriate focal length for the St\"{a}ckel potential at its present-day coordinates following \citet{Sanders2012}. Many authors have tested the accuracy of the St\"{a}ckel approximation for realistic Galactic potentials by comparing with integrated orbits, usually finding it to be better than $10\%$, and often about $2\%$  \citep[e.g. ][]{Binney2012,Sanders2012,2015ApJS..216...29B,Mackereth2018,Lane2022}. The approximation is best for disc-plane orbits, and tends to be worse for highly eccentric orbits, those which venture near the Galactic poles, or those in the bulge.

The three actions are often concisely displayed on a 2D plane with $J_{\phi}/J_\mathrm{tot}$ on the horizontal axis and $(J_{z}-J_{R})/J_\mathrm{tot}$ on the vertical axis \citep{2019MNRAS.488.1235M,Vasiliev2019,Lane2022}. Here, $J_\mathrm{tot} = J_{R} + \vert J_{\phi} \vert + J_{z}$ is the normalizing total action (note that $J_{R}$ and $J_{z}$ are manifestly positive). When actions are presented and scaled in this way, a characteristic diamond boundary defines the edges of the space, leading to its common name: the `action diamond'. The action diamond is useful in that it efficiently communicates the shape and orientation of an orbit, but can be limiting because its scale-free nature leads to degeneracies among orbits with similar shapes and orientations but different energies or radii. Nonetheless it is a valuable tool for studying associations among Galactic structures \citep{2019MNRAS.488.1235M,Vasiliev2019}, hence our choice in employing it here to study the dynamical similarities and differences between ejected GC stars.

Figure \ref{fig:actions} shows the action diamond for extra-tidal stars of all GCs in \cite{2018MNRAS.478.1520B}. For clarity, we only plot the actions of stars that escaped the cluster up to one orbital period ago. The action samples generally broaden when all stars are included. The distribution of individual GC actions (white points) is broadly uniform, which belies the isotropic nature of the galactic GC population. There is a notable enhancement, however, at the rightmost part of the diagram corresponding to prograde, disk-plane orbits. Figure \ref{fig:actions} also reveals that the distributions of extra-tidal stars are grouped rather tightly around their parent cluster, although there is certainly overlap between the samples in many cases. Hence, action variables can be a useful way to pinpoint specific parent GCs of extra-tidal stars in the field (see Section \ref{sec:puzzle} and Appendix \ref{sec:association} for a quantitative method that determines the relative probability of an observed extra-tidal field star being associated with a given Milky Way GC).

\subsection{Extra-tidal recoil binary properties} \label{sec:recoils}

The locations and actions of the escaped recoil binaries are extremely similar to those of the single extra-tidal stars presented in Figures \ref{fig:spatial} and \ref{fig:actions}, with the main differences being that (i) there are fewer binaries present and (ii) the distributions are more centrally concentrated with respect to the host clusters. Both of these differences can be attributed to the higher mass of the binary system compared to the single star, resulting in lower recoil velocity kicks received during the three-body interaction. Lower kick velocities for binaries ultimately result in fewer binary systems escaping the cluster. Furthermore, escaping binaries will have lower relative velocities compared to the host cluster than kicked single stars. The number of binaries that escape the cluster mainly depends on GC core density (closer encounters yield higher velocity kicks) and GC escape velocity (lower escape velocities yield more escaped binaries), where the escaped recoil binary fraction for each GC is shown in Figure \ref{fig:binary-frac}.

From Figure \ref{fig:binary-frac}, a correlation between GC core density $\rho_c$, escape velocity $v_{\rm esc}$, and escaped binary fraction $f_{\rm b, esc}$ is apparent. We opt to model the escaped binary fraction as a modified logistic function that depends on the (logarithm of) the core density and escape velocity:

 \begin{equation}
     f_{\rm b, esc}(\rho_c, v_{\rm esc}) = \frac{f_{\rm max}}{1 + \exp\left(-(a_0 + a_1 \log(\rho_c) + a_2 \ln(v_{\rm esc})\right)} + f_{\rm min}
     \label{eq:funcform}
 \end{equation}

 \begin{figure*}
    \centering
    \includegraphics[width=\textwidth]{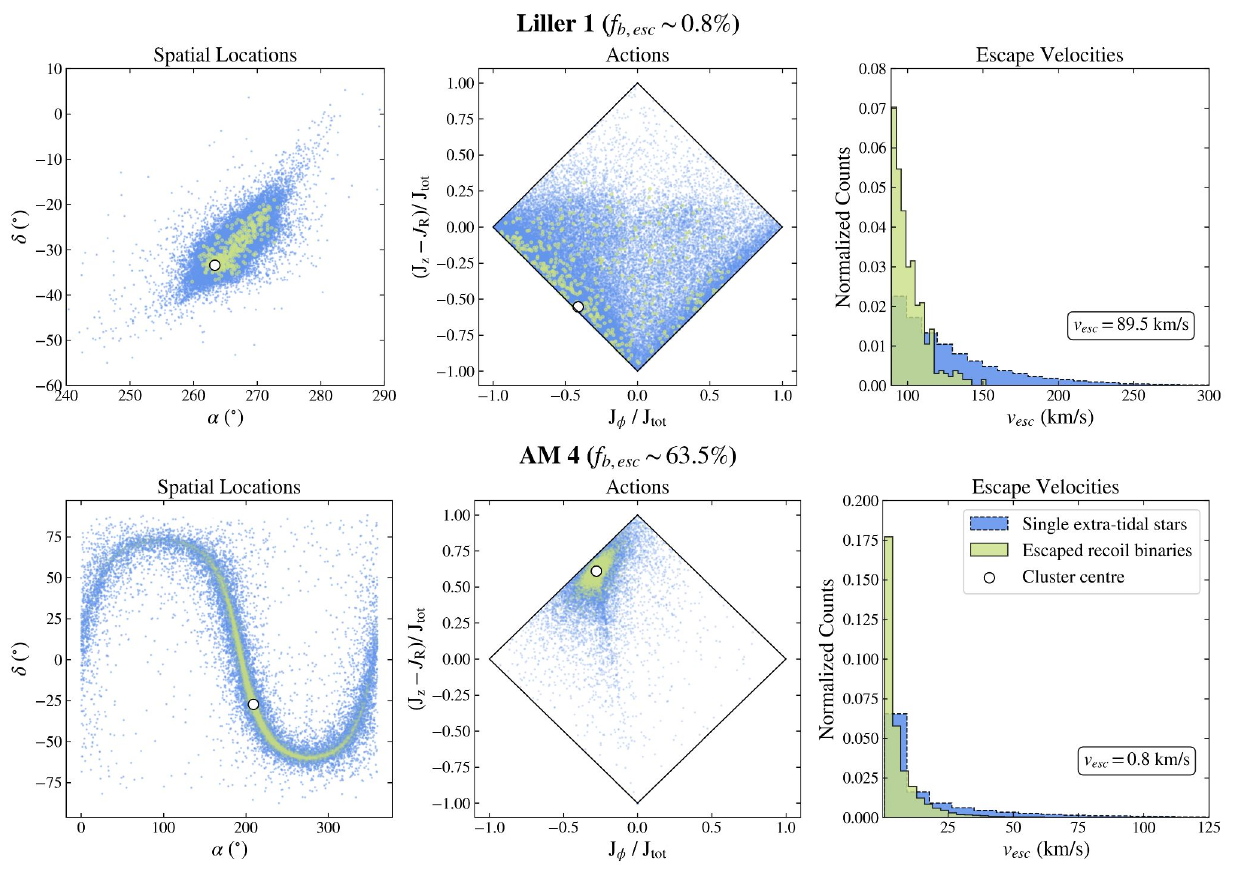}
    \caption{Extra-tidal stars and escaped recoil binaries of AM 4 and Liller 1 -- the GCs with the highest and lowest escaped recoil binary fractions, respectively ($0.8\%$ and $63.5\%$ in Figure \ref{fig:binary-frac}). For both clusters, the distributions of the singles and binaries have similar shapes, however the binaries (i) are more concentrated near the GC centre and (ii) have smaller escape velocities due to their heavier masses. AM 4 and Liller 1 have the lowest and highest escape velocities of any of the GCs in \citet{2018MNRAS.478.1520B}, and are used as illustrative examples to describe the similarities between the single and binary extra-tidal stars in the main parameter spaces examined. Note that  All stars and binaries were integrated in our baseline \texttt{MWPotential2014} Galactic potential model described in Section \ref{sec:pot-baseline}.}
    \label{fig:binaries}
\end{figure*}

\newpage

The free parameters $a_0$, $a_1$, and $a_2$ control the dependence of the escaped binary fraction on the core density and the escape velocity (the steepness and location of the curve), while $f_{\rm max}$ and $f_{\rm min}$ set the overall maximum and minimum allowed values. We define a log-likelihood for our fit assuming some intrinsic Gaussian scatter $s$ (an additional free parameter), giving us a total log-likelihood of our $n$ GCs:
\begin{equation}
    \ln \mathcal{L} = -\frac{1}{2} \left(\sum_{i=1}^{n} \frac{(f_{\rm b, esc}^i - f_{\rm b, esc}(\rho_c^i, v_{\rm esc}^i))^2}{s^2} + \ln(2\pi s^2) \right)
\end{equation}
We assume broad uniform priors over all six parameters ($\theta = \{a_0, a_1, a_2, f_{\rm max}, f_{\rm min}, s\}$), which are described in Appendix \ref{sec:model}. We sample from the posterior using \texttt{dynesty} \citep{2020MNRAS.493.3132S} \texttt{v2.1.3} with the default settings, but impose a final \texttt{dlogz} threshold of $10^{-3}$ to ensure denser sampling around the maximum likelihood solution. A corner plot summarizing the posterior and its uncertainties is shown in Appendix \ref{sec:model}, where our best-fit set of parameters, $\theta = \{a_0=3.39, a_1=-0.04, a_2=-3.37, f_{\rm max}=6.24 \times 10^{-1}, f_{\rm min}=1.28 \times 10^{-6}, s=7.94 \times 10^{-3}\}$, can be input into Equation \ref{eq:funcform} to compute $f_{\rm b, esc}$ for any GC with a given $\rho_c$ and  $v_{\rm esc}$.

\newpage
Based on our final set of posterior samples, we compute the posterior predictive for each $f_{\rm b, esc}(\rho_c, v_{\rm esc})$ by randomly resampling $\theta_j$ values from the final collection of samples. We compute the corresponding $f_{\rm b, esc}(\rho_c^i, v_{\rm esc}^i; \theta_j)$ predictions for those $\theta^j$ values, and add in scatter to the predictions based on the associated $s_j$ value. The posterior predictive residuals shown in the right-hand panel of Figure \ref{fig:binary-frac} are computed using the median and the 16th/84th percentiles derived from this process.

In Figure \ref{fig:binary-frac}, we also observe that the GC AM 4 exhibits the largest binary escape fraction, where $\sim 63\%$ of the binaries involved in three-body encounters producing single extra-tidal stars escape the cluster. AM 4 is also the cluster with the lowest escape velocity in the \cite{2018MNRAS.478.1520B} catalogue. Conversely, the GC Liller 1 has the smallest escape binary fraction (and highest escape velocity), where only $\sim 0.8\%$ of three-body encounter binaries escape the cluster. In Figure \ref{fig:binaries}, we compare the spatial and escape velocity distributions of the escaped binaries to the single extra-tidal stars of AM 4 and Liller 1. While escaped binaries of both clusters occupy a narrower distribution of orbital properties than escaped single stars, their properties are extremely comparable despite Liller 1 and AM 4 having the lowest and highest binary escape fractions.

\subsection{Extra-tidal stars in different Galactic potential models} \label{sec:potentialmodels}

In addition to simulating the production and orbital evolution of extra-tidal stars and binaries in the baseline \texttt{MWPotential2014} Galactic potential model, we also consider cases where host clusters and extra-tidal stars evolve in different potential models. As discussed in Section \ref{sec:models}, we consider a model that also has a bar and transient spiral arms (Bar+Transient Wave potential), a model with a heavier halo (\texttt{McMillan17}), non-spherical halos, and a model that accounts for the presence of the LMC and its resulting perturbations to the Milky Way. 

\begin{figure}
    \centering
    \includegraphics[width=0.5\textwidth]{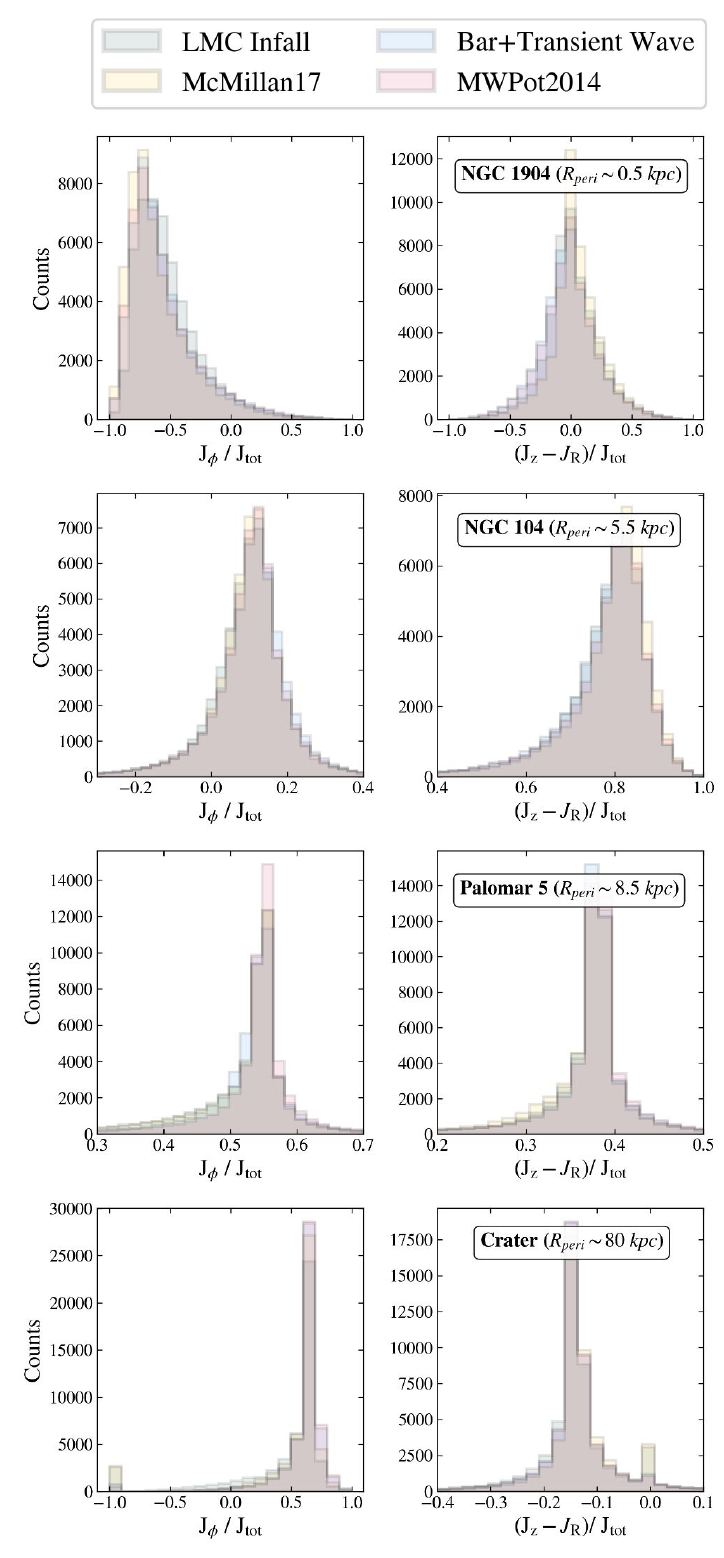}
    \caption{Action angle comparisons for extra-tidal stars from a disc (NGC 1904), bulge (NGC 104), inner halo (Palomar 5) and outer halo (Crater) cluster in four different Galactic potential models. \texttt{MWPotential2014} (Section \ref{sec:pot-baseline}) is represented in pink, Bar+Transient Wave (Section \ref{sec:pot-disc-transient}  in blue, \texttt{McMillan17} (Section \ref{sec:pot-mcmillan}) in yellow and a potential with an infalling LMC (Section \ref{sec:pot-lmc}) in green. Overall, there are very few differences between the extra-tidal star action distributions for each potential, where the minor offsets between the distributions are likely due to interactions with substructure in the Milky Way.}
    \label{fig:potential-comparison}
\end{figure}

To observe how choice of Galactic potential model affects the kinematics of extra-tidal stars, we compare the action angle coordinates for extra-tidal stars between our static and time-dependent potentials. To span the complete action diamond, we examine actions for clusters located in different characteristic parts of the Galaxy. Specifically, we examine action coordinates in each potential for a disc cluster (NGC 104), a bulge cluster (NGC 104), an inner halo cluster (Palomar 5) and an outer halo cluster (Crater) in Figure \ref{fig:potential-comparison}. 

Note that actions are always calculated in the baseline \texttt{MWPotential2014} Galactic potential using the St\"{a}ckel approximation in \texttt{galpy} \citep{2015ApJS..216...29B}, but the clusters have evolved in their own respective potentials. Hence, differences in the distributions of extra-tidal star actions are likely due to differences in the 6D-coordinates of the extra-tidal stars and not due to the assumed underlying potential. This approach is specifically necessary for models with time-dependent potentials as a calculation of extra-tidal star actions is not possible. This comparison is similar to assuming a potential like \texttt{MWPotential2014} to calculate the actions of observed stars, even though they have evolved in the true potential of the Milky Way.

For each cluster, there are only minor differences in actions between the potentials, indicating that model of the Galactic potential has little influence on the actions of the extra-tidal stars. The general slight offsets in actions between the models could partially be attributed to artificial effects like the St\"{a}ckel approximation in \texttt{galpy} \citep{2015ApJS..216...29B}. This occurs when a large gradient in the Galactic potential is present (e.g. in the inner regions of the Galaxy) and yields slightly different estimates of a cluster’s orbit depending on its orbital phase and the degree to which its orbit precesses every orbital period (see Appendix \ref{sec:Staeckel} for a full description of St\"{a}ckel effects). It is also possible that general interactions with various Galactic substructures could cause the orbits, and therefore orbit actions, of extra-tidal stars and their host GCs to change \citep{2020MNRAS.494.2268W, Garrow2020}. 

The most prominent difference among models is for the GC Crater, where a slight bi-modality in the action distributions is enhanced. This could indeed be an artifact, given that Crater lies towards the edge of the action diamond. However action differences in this cluster are perhaps not unexpected, as the LMC would influence outer halo clusters the most.

Note that each of the above potentials assume a perfect dark matter halo sphericity (i.e. $q=1.0$). However, when varying levels of halo oblateness as described in Section \ref{sec:nonspherical}, we once again find that the action distributions of the extra-tidal stars are comparable. This finding implies that the triaxiality of the Milky Way has little effect on the properties of extra-tidal stars produced by three-body encounters in GC cores, even for stars located far out in the Galactic halo (e.g. Crater).
 \begin{figure*}
    \centering
    \includegraphics[width=\textwidth]{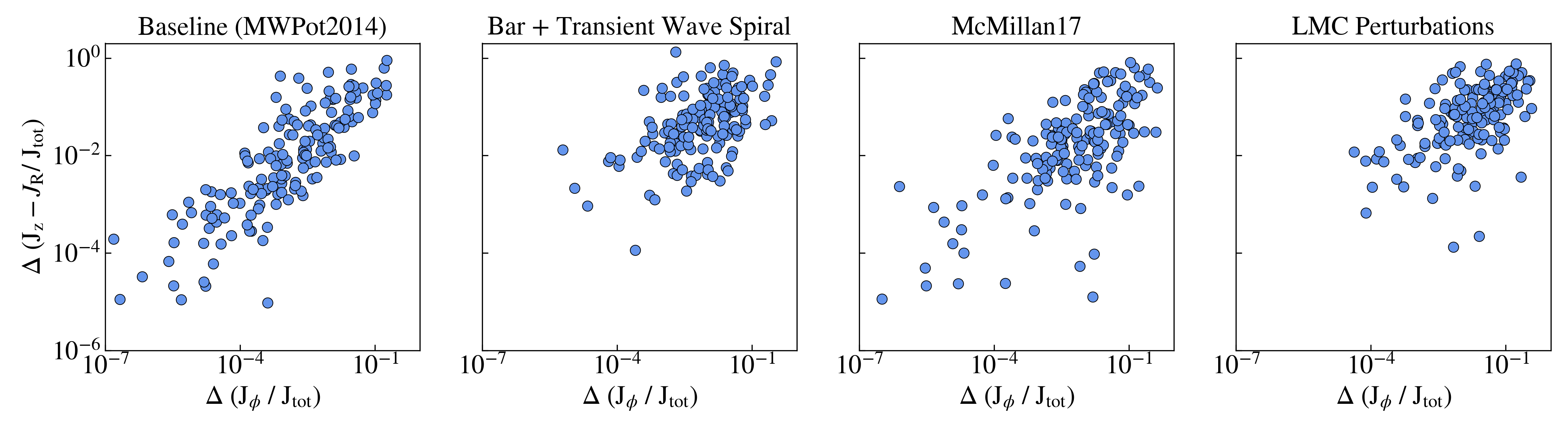}
    \caption{Absolute magnitudes of the changes in $J_{\phi}/J_{tot}$ and $(J_{R}-J_{z})/J_{tot}$  when calculated using a GC's current orbital coordinates and when calculated using its orbital coordinates five orbital periods ago. Clusters have had their orbits integrated in either the baseline \texttt{MWPotential2014} Galactic potential (first panel), the Bar+Transient Wave potential (second panel), the \texttt{McMillan17} Galactic potential (third panel), or the \texttt{MWPotential2014} in the presence of the Large Magellanic Cloud (fourth panel) (see Section \ref{sec:models} for full descriptions of each model). In the static baseline \texttt{MWPotential2014} Galactic potential, a GC's actions will artificially evolve due to our use of the St\"{a}ckel approximation to calculate cluster actions. In time-dependent potentials, cluster actions will evolve primarily due to both interactions with substructure as well as our use of the St\"{a}ckel approximation.}
    \label{fig:actionevolve}
\end{figure*}

\newpage

\subsection{The time (in)dependence of extra-tidal actions}\label{sec:time}

\begin{figure*}
    \centering
    \includegraphics[width=0.9\textwidth]{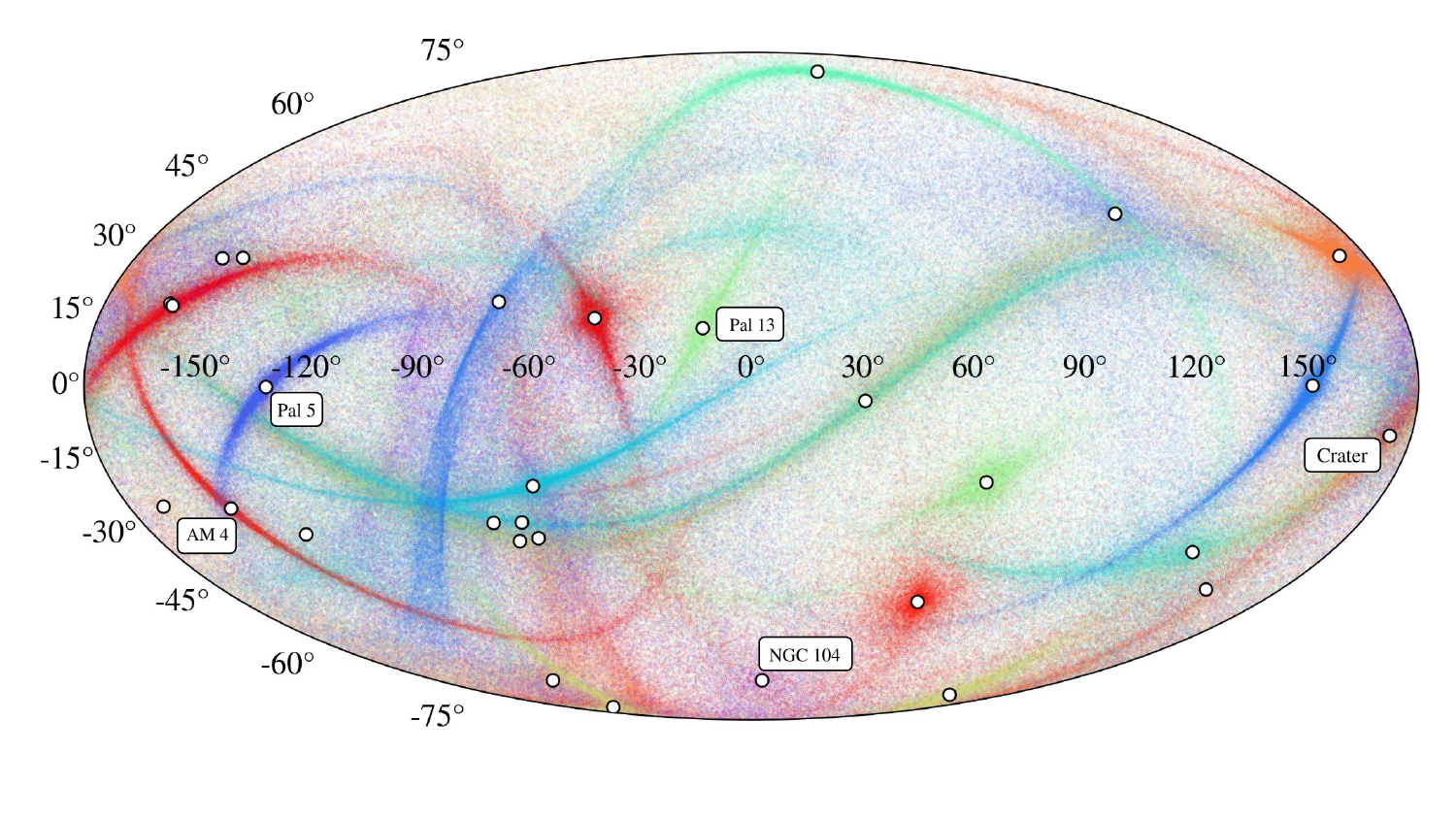}
    \caption{A Mollweide projection for extra-tidal mock star distributions of GCs with pericentre radii $R_{\text{peri}} > 5$kpc in \citet{2018MNRAS.478.1520B}. Clusters specifically discussed throughout this work are labelled. All-sky stream-like behaviour is prominent for these clusters, indicating that dynamically ejected stars from GC cores could populate observed stellar streams or tidal tails, which are currently only believed to be produced through tidal stripping.}
    \label{fig:stellarstreams-mollweide}
\end{figure*}

While the specific rate of three-body encounters in a GC’s core depends on parameters like cluster core density and mass, extra-tidal stars will be ejected at random times throughout a GC’s orbit around the Galaxy. To investigate how cluster actions evolve as a function of time, we compare the present-day actions of each host GC to its actions based on its orbital coordinates five orbital periods in the past in Figure \ref{fig:actionevolve}.

In \textit{time-independent} potentials (e.g. \texttt{MWPotential2014} and \texttt{McMillan17}), any evolution in cluster and extra-tidal actions is artificial and likely due to our use of the St\"{a}ckel approximation, as discussed in Appendix \ref{sec:Staeckel}. In most cases however, the shift is small, as $81\%$ and $70\%$ of clusters show shifts of less than 0.1 along both axes in the \texttt{MWPotential2014} potential (left-most panel of Figure \ref{fig:actionevolve}) and \texttt{McMillan17} potential (third panel of Figure \ref{fig:actionevolve}) respectively. If we compare the \texttt{Corespray} action distributions of the recent extra-tidal escapers to those that escaped early on in the \texttt{MWPotential2014} Galactic potential model, we see that all but four clusters have the means of their distributions shift by less than 0.1. The four outliers -- ESO 452-SC11, NGC 6380, NGC 6453 and NGC 6558 -- all have $R_{\text{peri}} <0.5$kpc. It is therefore not surprising that the St\"{a}ckel approximation predicts different actions for the cluster as a function of time due to the gradient in the potential throughout this region, despite the potential being static.

In \textit{time-dependent} potential models (e.g. Bar+Transient Wave Spiral and infall due to the LMC), a cluster's actions will truly evolve due to interactions with substructure as opposed to the artificial evolution we see in the static potentials. In the Bar+Transient Wave potential (second panel of Figure \ref{fig:actionevolve}), the change is still minimal with $75\%$ of clusters experiencing shifts in $(J_{R}-J_{z})/J_{tot}$ and $J_{\phi}/J_{tot}$ less than 0.1 due to few clusters undergoing strong interactions with the Galactic disc. In fact, with most clusters orbiting in the halo, the presence of the LMC results in the largest evolution of cluster actions as only $57\%$ of clusters in this potential experience a shift along both axis less than 0.1 (right-most panel of Figure \ref{fig:actionevolve}). Hence for most clusters, any evolution in action space due to time-dependent substructure is comparable to how a cluster's calculated actions change with orbital phase when using the St\"{a}ckel approximation to calculate actions.

When comparing the GEMS catalogue to observations (e.g. when using Appendix \ref{sec:association} to associate field stars with individual GCs), it is important that actions are calculated in a self-consistent manner. If actions are calculated with the same method and in the same potential, the distributions given by \texttt{Corespray} simulations that use the baseline \texttt{MWPotential2014} Galactic potential model can be used to search for extra-tidal stars that (i) are close to their host cluster and (ii) escaped recently. For extra-tidal stars that escaped long ago, the effects of spiral arms or the LMC should be considered depending on whether a low-latitude or outer region cluster is being studied.

\section{Discussion}\label{sec:discussion}

\subsection{Core stars and binaries can contaminate stellar streams} \label{sec:streams}

\subsubsection{The tidal tails of Palomar 13}

Stellar streams are elongated distributions of escaped stars from GCs that are mainly believed to be the result of tidal stripping by the host Galaxy. In this scenario, mass segregation causes low mass stars in the GC to migrate to the outskirts, where tidal forces from the host galaxy gradually pull the stars away. Stars then escape through a cluster's Lagrange points with velocities that are near-zero relative to the cluster itself. This process happens while the GC orbits the Galaxy, causing stars to form long, stretched out associations that lead and trail the cluster \citep[e.g.][for an observational review of stellar streams in the Galactic halo]{2016ASSL..420...87G}. 

As previously illustrated in Figure \ref{fig:spatial}, extra-tidal star distributions for GCs with $R_{\text{peri}} > 5$kpc in \cite{2018MNRAS.478.1520B} can exhibit stream-like behaviour. We further investigate the structure of these distributions using an all-sky Mollweide projection in Figure \ref{fig:stellarstreams-mollweide}. Here, it is evident that extra-tidal stars from halo GCs are primarily found in stellar streams, which are found all throughout the Milky Way. As discussed in Section \ref{sec:spatial}, these stream-like properties often occur when a cluster has a low core escape velocity, such that three-body interactions that yield weak kicks result in the single star escaping the host cluster with a near-zero velocity relative to the cluster itself. Even though the kick is occurring in a random direction and the star isn't necessarily escaping through the cluster's Lagrange points like a tidally stripped star, the escaping star will still follow an orbit that is comparable to its progenitor cluster. Hence, the escaped stars populate streams that tail the cluster along its orbit of the Galaxy.

\newpage 

\begin{figure}
    \centering
    \includegraphics[width=\columnwidth]{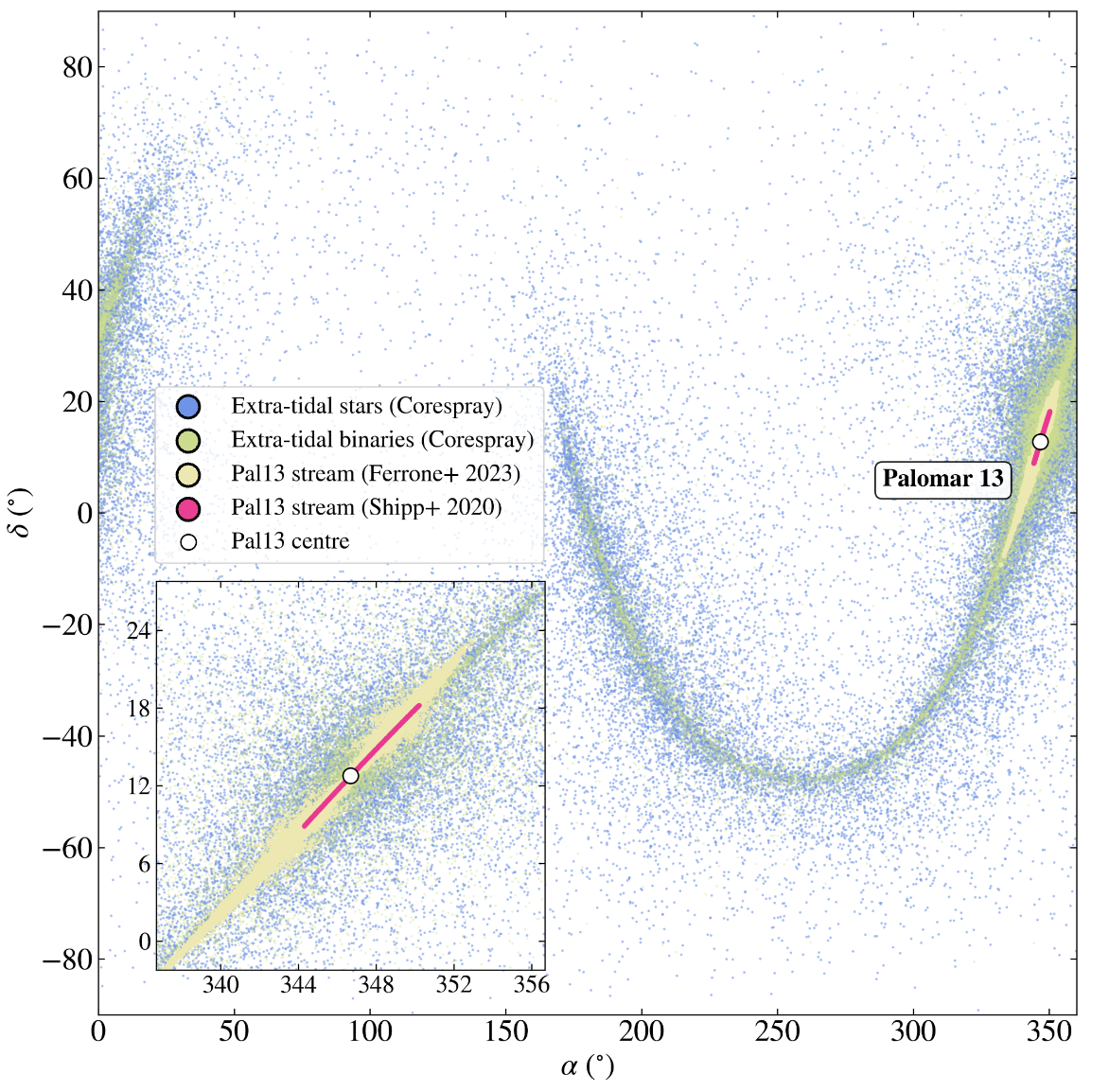}
    \caption{A comparison between extra-tidal stars, binaries and  stellar stream stars of the GC Palomar 13. \texttt{Corespray} extra-tidal mock stars and binaries from the core of Palomar 13 are indicated in blue and green, respectively. \citet{2023A&A...673A..44F} simulate tidally-stripped stars for Palomar 13, which are indicated in yellow. Observed Palomar 13 stellar stream data from \citet{2020AJ....160..244S} is plotted in pink using \texttt{galstreams} \citep{2023MNRAS.520.5225M}. The centre of Palomar 13 is indicated with a white circle. Stream behaviour is visible for both the extra-tidal single stars and binaries, which are also aligned with both the simulated tidal tails and the observed stellar stream stars. Note that ejection is along the direction of the stream in \citet{2023A&A...673A..44F}, which results in a lack of stars around the cluster centre. Ultimately, this comparison indicates that GC core stars and binaries could contaminate the Palomar 13 stellar stream, which up to this point, is believed to only contain stars from tidal stripping.}
    \label{fig:stellarstreams}
\end{figure}

To explore this behaviour in more detail, in Figure \ref{fig:stellarstreams} we plot the locations of extra-tidal tidal stars and binaries from the GC Palomar 13 -- a cluster that was recently shown to have extended tidal tails \footnote{See \citet{Piatti2020} for a complete list of GCs with observed tidal tails.} \citep{2020AJ....160..244S}. From our \texttt{Corespray} simulation, it is immediately clear that both the single and binary extra-tidal stars of Palomar 13 exhibit stream-like behaviour. To further explore this statement, we compare our Palomar 13 \texttt{Corespray} stars to the \texttt{Pal13-S20} stream in \texttt{galstreams} \citep{2023MNRAS.520.5225M} utilizing observed data from \cite{2020AJ....160..244S}. From Figure \ref{fig:stellarstreams}, we see that the locations of the extra-tidal stars and binaries ejected from the core of Palomar 13 are consistent with the observed stellar stream stars. This overlap allows us to conclude that the tidal tails of Palomar 13 could be contaminated with stars and binaries from the cluster's core. \textit{Importantly, this finding challenges the current belief that GC tidal tails or stellar streams are purely the result of tidal stripping, indicating that dynamical interactions such as three-body encounters could influence their overall composition.}

\begin{figure*}
    \centering
    \includegraphics[width=\textwidth]{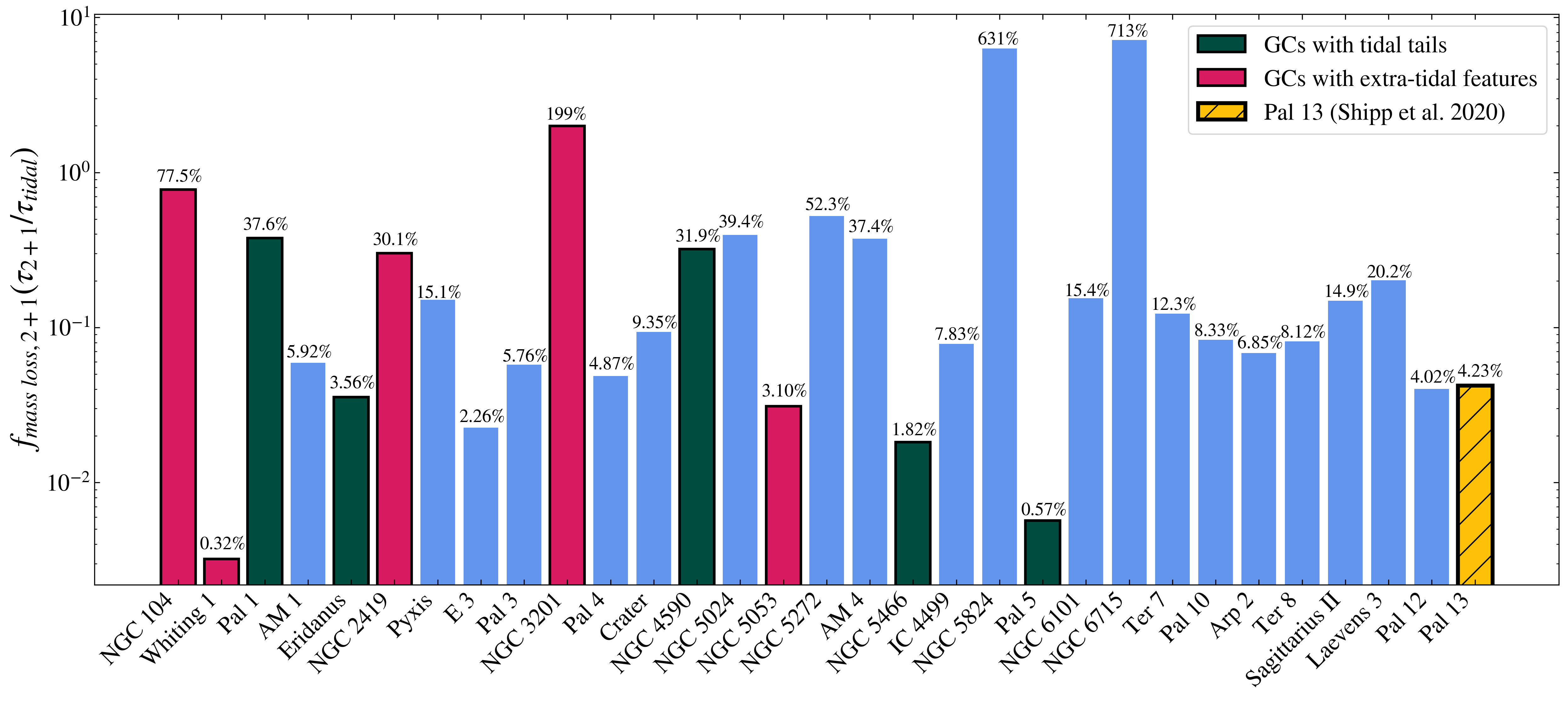}
    \caption{The cluster mass loss fraction due to three-body encounters compared to tidal stripping, $f_{\text{mass loss}, 2+1}$. For all 31 GCs in our sample with $R_{\text{peri}} > 5$kpc, we compute $f_{\text{mass loss}, 2+1} = \tau_{2+1} / \tau_{\text{tidal}}$ from Equations \ref{eq:totalmassloss} and \ref{eq:2+1rate}. Clusters with known tidal tails or extra-tidal features presented in \citet{Piatti2020} are highlighted in pink and green, respectively. Palomar 13 -- our case study cluster -- is highlighted in yellow. It is evident that mass loss due to three-body encounters is non-negligible for some clusters, providing further evidence that stellar streams could be contaminated with core stars or binaries. Importantly, $f_{\text{mass loss}, 2+1}$ represents an \textit{approximate} mass loss rate, as not all three-body encounters will result in stellar escape from a cluster. Note that $f_{\text{mass loss}, 2+1}$ is displayed in log-scale.}
    \label{fig:massloss}
\end{figure*}

Stellar streams are also believed to be kinematically ``cold'', as stars escape a cluster with velocities close to the GC's escape speed due to the constant pull of Galactic tidal forces. For three-body core encounters, however, the overall kick velocity distribution is governed by the core velocity dispersion of the cluster. The GC's escape speed, on the other hand, just sets a boundary for what fraction of the kick distribution will actually escape. While core interactions yielding weak velocity kicks will result in stars escaping the host cluster with near-zero velocities relative to the cluster itself (comparable to tidal stripping), dynamical interactions can eject stars with a variety of velocities, often much higher than what would occur during tidal stripping. \textit{Thus, we also conclude it is possible that a stream contaminated by core stars and binaries could appear dynamically hotter and thicker than if it was just populated by tidally stripped stars.} This finding has implications for methods that attempt to use stream properties to constrain the stream's progenitor \citep{1998ApJ...495..297J, 2019MNRAS.485.5929W,2021ApJ...911L..32G} or infer interactions with dark matter substructure \citep{2017MNRAS.466..628B,2019ApJ...880...38B}. For clusters with high three-body interaction rates, kicked core stars with stream-like orbits may also appear as diffuse stellar structure around the stream, similar to the observed cocoon around GD-1 \citep{2019ApJ...881..106M}. 

\subsubsection{Stream contamination fractions from three-body encounters}

While direct $N$-body simulations with realistic binary fractions would be required to estimate the exact amount of contamination and the degree to which core stars might make a stream appear dynamically hotter, we can estimate the fraction of core stars from three-body encounters that contribute to the cluster mass loss rate due to tidal stripping. Equation \ref{eq:totalmassloss} represents the total mass loss rate of the cluster ($\tau_{\text{tidal}})$, where $M$ is the cluster mass from \cite{2018MNRAS.478.1520B} and $T_{\text{diss}}$ is the remaining cluster lifetime from \cite{2003MNRAS.340..227B}. Importantly, we treat this total mass loss rate to be approximately representative of the mass loss rate due to tidal stripping, since this is the dominant mass loss channel of evolved clusters in the present-day \citep[e.g.][]{2003gmbp.book.....H}. 

\begin{equation}
    \tau_{\text{tidal}} = \dfrac{M} {T_{\text{diss}}} [M_{\odot}/\text{Myr}]
    \label{eq:totalmassloss}
\end{equation}

The three-body interaction rate due to GC core three-body encounters is computed via Equation A10 in \cite{2011MNRAS.410.2370L}, where $f_{b}$ is the binary fraction, $f_{t}$ is the triple fraction, $r_{c}$ is the core radius, $n_{0}$ is the mean stellar density in the core \citep[computed from $\rho_{c}$, $M/L$ and $\langle m \rangle$ via Equation 13 in][]{2011MNRAS.410.2370L}, $v_{rms}$ is the velocity dispersion, $\langle m \rangle $
is the mean stellar mass and $a_{b}$ is the semi-major axis of the binary. Note that to get a mass loss \textit{rate} ($\tau_{2+1}$), we assume an 100\% stellar escape fraction and an average stellar escaper mass of $\langle m \rangle = 0.5 M_{\odot}$ and divide this by Equation 13 in \cite{{2011MNRAS.410.2370L}}, yielding Equation \ref{eq:2+1rate}.

\begin{multline}
    \tau_{2+1} = \dfrac{0.5 M_{\odot}}{34} \times (1 - f_{b} - f_{t}) \left(\dfrac{1 \text{pc}}{r_{c}} \right)^{-3} \left(\dfrac{10^{3} pc^{-3}}{n_{0}}\right)^{-2}  \\ \left(\dfrac{v_{rms}}{5 \text{km s}^{-1}}\right)^{-1} \left(\dfrac{0.5 M_{\odot}}{\langle m \rangle}\right)^{-1} \left(\dfrac{1\text{au}}{a_{b}}\right)^{-1} [M_{\odot}/\text{Myr}]
    \label{eq:2+1rate}
\end{multline}

To solve for $\tau_{2+1}$, we adopt values of $f_{b}=0.1$ and $f_{t}=0.01$, where we acknowledge that in reality, binary fraction decreases with cluster density \citep{2012A&A...540A..16M}. All GC structural parameters \citep[e.g. $r_{c}$, $v_{rms}$, $\rho_{c}$ and $M/L$ are taken from][]{2018MNRAS.478.1520B}. We also assume $\langle m \rangle = 0.5$ and $a_{b}=a_{hs}$, where $a_{hs}$ corresponds to the hard-soft limit for binaries.

The mass loss fraction due to three-body encounters can be estimated by dividing Equation \ref{eq:2+1rate} by Equation \ref{eq:totalmassloss}, where $f_{\text{mass loss}, 2+1} = \tau_{2+1} / \tau_{\text{tidal}}$. Figure \ref{fig:massloss} highlights $f_{\text{mass loss}, 2+1}$ for all GCs with $R_{\text{peri}} > 5$kpc in our sample, as these are the clusters that typically exhibit stream-like behaviour in Figure \ref{fig:spatial}. Clusters with tidal tails or extra-tidal features presented in \cite{Piatti2020} are highlighted in pink and green, respectively. Our case study cluster Palomar 13 is highlighted in yellow.

From Figure \ref{fig:massloss}, it is clear that for many clusters, three-body encounters in GC cores can contribute a non-negligible amount to the mass loss rate due to tidal stripping. This holds true for Palomar 13, where we find that $4.23\%$ of the mass loss rate due to stripping could actually come from three-body encounters. Hence, it is indeed possible that a fraction of real stars in these streams could be extra-tidal single stars or binaries from the cluster cores.While such contamination would also make it more difficult to constrain a cluster's dynamical history from variation in the mass function along the stream \citep{Webb2022}, it may also provide further insight into the time evolution of a GC's core. 

It is of course important to note that $f_{\text{mass loss}, 2+1}$ is a simplified approximation. On one hand, not all three-body encounters will result in stellar escape from a cluster. However, mass segregation would imply that core stars are more likely to be brighter than most low-mass tidally stripped stars that have segregated to a cluster's outer regions. Additionally, only a fraction of the stars that escaped via tidal stripping would be bright enough to observe given magnitude limits. Hence, the contamination fraction of \textit{observed} core stars in stellar streams could be higher than our estimates here. A detailed quantitative follow-up of stellar stream contamination from GC core stars, binaries and even compact objects will be available in Grondin \& Webb 2024, in preparation.

\subsection{Identifying an extra-tidal star's progenitor cluster} \label{sec:puzzle}

In order to quantify and illustrate how unique a given region of the action diamond parameter space is to extra-tidal stars from a single GC, we first estimate the probability distribution function $\mathbb{P}((J_z - J_R)/J_{tot}$, $J_{\phi}/J_{tot})$ of each cluster's mock extra-tidal stars. A Gaussian kernel density estimate of each distribution is determined from the extra-tidal mock stars \footnote{See the \texttt{gaussian\_kde} function in the \texttt{scipy} Python package \citep{2020SciPy-NMeth}.}, after which the probability distribution function of each cluster is evaluated for a grid of points that spans the action diamond. At each location within the action diamond, we evaluate $\mathbb{P}$ for each cluster and determine the cluster that yields the maximum $\mathbb{P}_{max}$. Furthermore, for the cluster with the highest evaluation of $\mathbb{P}$ at a given location, we also calculate the logarithm of the odds ratio between that cluster and each of the other clusters in our dataset $\log \mathbb{P}_{max}/ \mathbb{P}_{remaining}$. Hence, the relative probability that an extra-tidal star with a given set of actions escaped from one GC to another can be determined, where the cluster with the highest probability will represent the extra-tidal star's most likely origin.

\begin{figure*}
    \centering
    \includegraphics[width=0.97\textwidth]{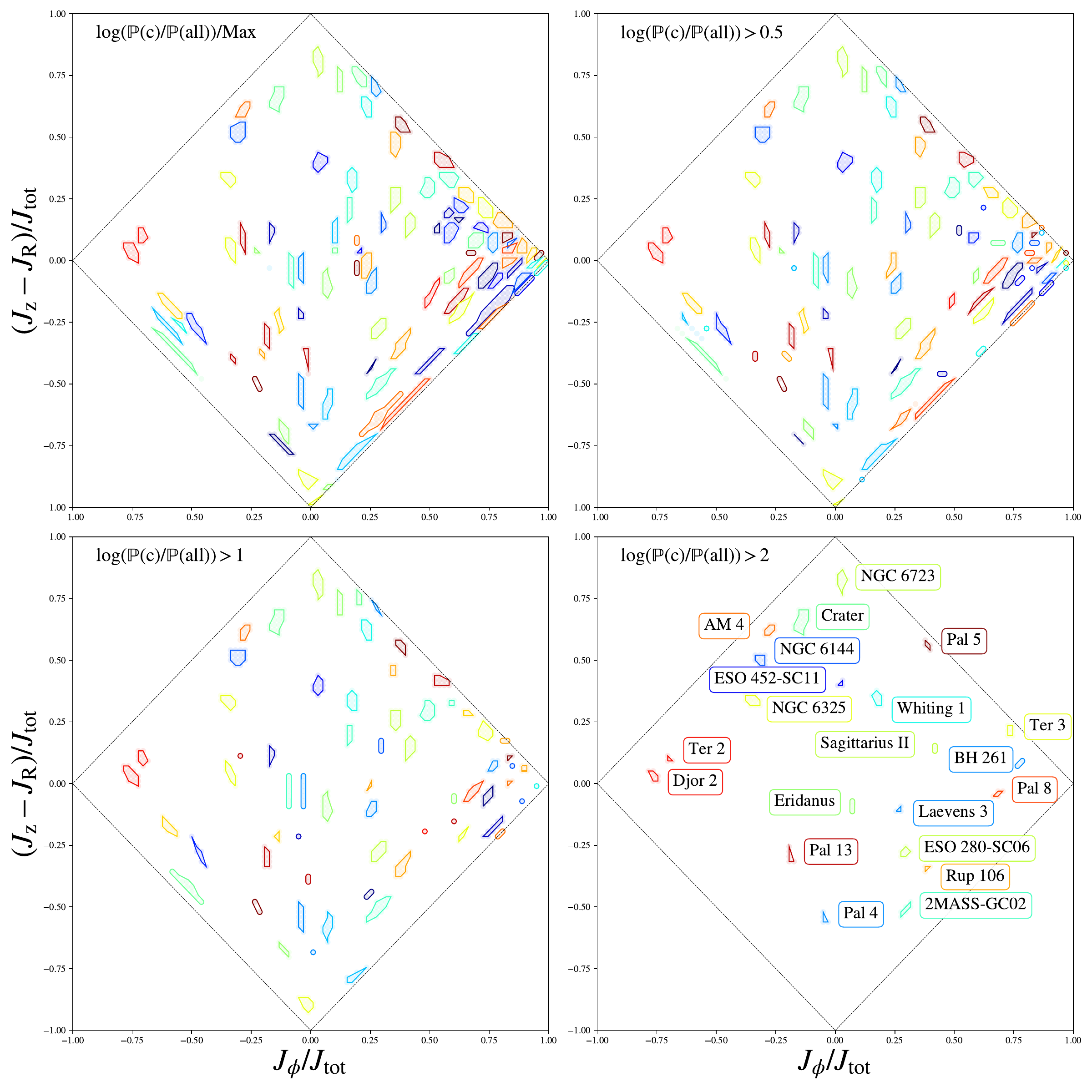}
    \caption{Action diamond maps that are colour coded by the cluster with the probability distribution function that yields the highest evaluation of $\mathbb{P}$ at each location where there are more than 500 \texttt{Corespray} stars from a single cluster. Lines are drawn between edgepoints using the \texttt{alphashape} Python package (Bellock, K.E., \href{https://alphashape.readthedocs.io/en/latest/readme.html}{https://alphashape.readthedocs.io/en/latest/readme.html}). Using the \citet{kassraftery} metric, we only show regions where the logarithm of the ratio in $\mathbb{P}$ of the most likely cluster and $\mathbb{P}$ of all other clusters is greater than 0.5 (top-right), 1 (bottom left), and 2 (bottom right). While the action diamond can be divided into regions that are more likely associated with a given cluster, the overlap between mock extra-tidal star distribution functions results in a narrowing of the parameter space from which one can strongly or decisively match a field star to a host cluster.}
    \label{fig:puzzle}
\end{figure*}

Figure \ref{fig:puzzle} illustrates four maps of the action diamond parameter space, colour coded by the cluster that yields $\mathbb{P}_{max}$ at a given location. Outlines are drawn to highlight connected regions where the  $(J_z - J_R)/J_{tot}$ and $J_{\phi}/J_{tot}$ values of mock extra-tidal stars indicate the stars are likely to share a common origin. In the top-left panel of Figure \ref{fig:puzzle}, each location is mapped to the cluster that yields the highest evaluation of $\mathbb{P}$ regardless of how this cluster's probability distribution compares to other clusters. We also require that at least 500 \texttt{Corespray} stars from the cluster that yield $\mathbb{P}_{max}$ fall within a given data point to include it in Figure \ref{fig:puzzle}. Therefore, while this panel clearly demonstrates how the action diamond space can be broken up into regions that are more strongly associated with a given cluster, it does not take into consideration the fact that the distribution functions of other clusters might yield similar estimates of $\mathbb{P}$.

To identify regions of the action diamond parameter space where an extra-tidal star's most likely origin is significantly more probable than any other cluster, we focus on regions where $\log \mathbb{P}_{max}/ \mathbb{P}_{remaining}$ satisfies specific criteria. The maps in the top-right, bottom-left, and bottom-right panels of Figure \ref{fig:puzzle} are again colour coded by the cluster that yields $\mathbb{P}_{max}$ at a given location, but only show cases where $\log \mathbb{P}_{max}/ \mathbb{P}_{remaining}$ is greater than 0.5, 1, and 2 respectively. These thresholds correspond to cases where the estimate of an extral-tidal star's most likely origin are significant, strong, and decisive, using the  \cite{kassraftery} metric. If we first consider the top-right panel of Figure \ref{fig:puzzle}, all whitespace can be considered as regions where the mock extra tidal star distributions of at least two clusters overlaps too much to strongly conclude that a star at the location in the action diamond is from one specific cluster. Regions that retain their colour coding demonstrate locations where an estimate of an extra-tidal star's origin would be significant.

Taking into consideration the bottom two panels in Figure \ref{fig:puzzle}, we see that the parameter space over which one can strongly or decisively state that an extra-tidal star comes from a specific cluster becomes smaller and smaller compared to the $\log \mathbb{P}_{max}/ \mathbb{P}_{remaining} > 0.5$ case. However, there remain 21 clusters that occupy truly unique regions of the action diamond parameter space, such that a star observed in that region would most likely come from that cluster compared to any other. The 21 clusters labeled in the bottom-right corner of Figure \ref{fig:puzzle} represent the most ideal GC candidates for kinematically searching for extra tidal stars. Appendix \ref{sec:association} outlines a Python code that (i) identifies the individual GC with the highest probability of association and (ii) computes the logarithm of the odds ratio between the first and second most likely progenitor clusters for any field star or binary.

It is important to note that these probability distribution functions do not represent the probability that a star was ejected from a given GC. Such a probability cannot be known without knowledge (or an estimate of) the dissolved star cluster population of the Milky Way and the orbital distribution of field stars in a given region. However, as described in Section \ref{sec:intro}, quantitatively associating stars with progenitor clusters is often done by incorporating chemistry along with dynamics (`chemo-dynamical tagging'). If the chemical properties of an observed field star are also comparable to those of a GC that yields a large value of $\mathbb{P}$, it is stronger evidence that the field star was once located in the candidate cluster. A full Milky Way scale association of field stars to progenitor GCs could thus be achieved by applying the methods in \cite{2023MNRAS.518.4249G} to clusters in the GEMS catalogue that have chemical data, for example. 

The GEMS catalogue also allows for a more rigorous estimate of $\log \mathbb{P}_{max}/ \mathbb{P}_{remaining}$ than presented in Figure \ref{fig:puzzle}. Calculating the probability distribution function from a four-dimensional Gaussian kernel density estimate of extra-tidal star distributions in $J_r$, $J_{\phi}$, $J_z$, and orbital energy space would lift some of the degeneracies that exist in the action diamond parameter space. These four dimensions have been used to link Galactic GCs to specific accretion events in the Milky Way's history \citep{2019MNRAS.488.1235M}. The probability of association could also be improved by incorporating the expected number of ejected stars from each cluster given their current mass loss rates, or an evolutionary track, into the calculation. 

\section{Conclusions and Implications}\label{sec:conclusions}

In this study, we present the GEMS (Globular cluster Extra-tidal Mock Star) catalogue; a collection of simulated extra-tidal stars and binaries created from three-body encounters in GC cores. Extra-tidal stars and binaries of all 159 GCs listed in \cite{2018MNRAS.478.1520B} are simulated using the \texttt{Corespray} software \citep{2023MNRAS.518.4249G}, with stellar orbits integrated in seven different Galactic potential models. The main results and implications are summarized below.
\newpage
\begin{enumerate}

    \item Extra-tidal stars and binaries of individual GCs generally occupy unique regions in both (i) location in the Galaxy (Figure \ref{fig:spatial}) and (ii) the action diamond (Figure \ref{fig:actions}). Field stars and binaries can be matched with possible progenitor GCs using the kinematics presented in the GEMS catalogue (Section \ref{sec:puzzle}). Coupled with additional chemo-dynamical information, this tool can provide associations with corresponding probabilities of any given field star being associated with any Galactic GC. Furthermore, there are 21 clusters that occupy extremely unique regions of the action diamond (Figure \ref{fig:puzzle}), allowing for decisive associations of field stars with these clusters via the \cite{kassraftery} metric (Appendix \ref{sec:association}).

    \item We present a modified logistic function model for the escaped recoil binary fraction of any Galactic GC as a function of core density and cluster escape velocity in Equation \ref{eq:funcform}. The best-fit model parameters are determined from the posterior predicative values calculated with the \texttt{dynesty} \citep{2020MNRAS.493.3132S} nested sampling code.

    \item Extra-tidal star characteristics are largely independent of the choice of Galactic potential model, where large-scale disc structure, perturbations from the LMC and mass and sphericity of the dark matter halo all appear to mostly be negligible (Section \ref{sec:potentialmodels} and Figure \ref{fig:potential-comparison}). For the distributions of our \texttt{Corespray} simulated extra-tidal stars, slight differences in action coordinates are found between recent and old escapers for inner region clusters with $R_{\text{peri}} <0.5$kpc. However, this may be due to our use of the St\"{a}ckel approximation when calculating actions (Section \ref{sec:time} and Appendix \ref{sec:Staeckel}). 
    
    \item Extra-tidal stars and binaries produced from three-body encounters in GC cores could contaminate stellar streams (Section \ref{sec:streams}, Figures \ref{fig:stellarstreams} and \ref{fig:massloss}). Contamination from core stars that escape with relative velocities higher than tidally stripped stars may cause streams to appear hotter. Since it is mainly believed that GC stellar streams are the products of tidal stripping (and thus contain low-mass stars), this novel result unveils possibilities for constraining stream progenitors or even explaining diffuse structure around observed streams (e.g. the observed cocoon around GD-1).
    
\end{enumerate}

\subsection{Future work}

The GEMS catalogue provides a large sample of extra-tidal stars and binaries that can be used to probe cluster dynamics and evolution at a population-level scale. A few immediate applications of this catalogue and \texttt{Corespray} are listed below:

\begin{enumerate}
    \item The rate at which three-body interactions occur is highly dependent on the number, core density and velocity of stars in the core of the cluster itself \citep{2008gady.book.....B} and the properties of its binary star population \citep{2011MNRAS.410.2370L}. Taking these factors into consideration, one could use \texttt{Corespray} or sample the GEMS catalogue to generate realistic realizations of extra-tidal star distributions based on the expected number of extra-tidal stars ejected from each GC to determine the best GCs to observationally find extra-tidal stars. These identified stars could then be matched to individual GCs using the framework presented in Section \ref{sec:puzzle} and Appendix \ref{sec:association}.
     \item In this study, we find that the spatial and action coordinate distributions of extra-tidal stars and binaries generally broaden as binaries escape at earlier times in the past. However, in clusters with high three-body interaction rates, one could use \texttt{Corespray} to probe cluster binary fraction as a function of time. This approach would allow for insights into the fraction of binaries that exist in a cluster at a given time, providing useful constraints for compact object interactions and high-velocity star production in the cores of GCs. Such an approach would also have to factor in the evolution of the binary populations binding energy distribution as well \citep{Leigh2022}.
\end{enumerate}

Ultimately, combining extra-tidal star data from the GEMS catalogue with additional chemo-dynamical information (e.g. from APOGEE and \textit{Gaia}) will allow field stars to be associated with individual Galactic GCs. As more extra-tidal stars are associated with clusters, we can continue to better our understanding of GC dynamics and evolution and begin to tackle big open questions like cluster formation too. 

\section*{Acknowledgements}
The authors thank Jo Bovy, Jason Hunt and Ting Li who provided important context, insight and references for the various Galactic potential models used in this study. The authors also acknowledge Fraser Evans and Phil Van-Lane for providing thorough feedback on the manuscript itself. Advice from Samantha Berek, Victor Chan, Maria Drout, Mairead Heiger, Marten van Kerkwijk and Claire Ye also greatly improved this work. SMG acknowledges the support of the Natural Sciences and Engineering Research Council of Canada (NSERC) and is partially funded through a NSERC Postgraduate Scholarship -- Doctoral. SMG also recognizes current funding from a Walter C. Sumner Memorial Fellowship and an Ontario Graduate Scholarship (2022-2023). SMG wishes to thank Debbie, Dianne and Tom Postnikoff for meaningful support and atvar during the preparation of this manuscript. JJW acknowledges financial support from NSERC (funding reference number RGPIN-2020-04712) and an Ontario Early Researcher Award (ER16-12-061). JSS would like to acknowledge support from the Natural Sciences and Engineering Research Council of Canada (NSERC) funding reference \#RGPIN-2023-04849. The Dunlap Institute is funded through an endowment established by the David Dunlap family and the University of Toronto. NWCL gratefully acknowledges the generous support of a Fondecyt Regular grant 1230082, as well as support from Millenium Nucleus NCN19\_058 (TITANs) and funding via the BASAL Centro de Excelencia en Astrofisica y Tecnologias Afines (CATA) grant PFB-06/2007.  NWCL also thanks support from ANID BASAL project ACE210002 and ANID BASAL projects ACE210002 and FB210003.

\section*{Data Availability}

The GEMS catalogue is available online and can be downloaded at \href{https://zenodo.org/record/8436703}{https://zenodo.org/record/8436703}. There are two corresponding \texttt{csv} files containing data for (i) single extra-tidal stars and (ii) escaped recoil binaries, where orbits are integrated in our baseline \texttt{MWPotential2014} potential. A description of each column is provided in a supplementary \texttt{README.txt} file. All individual GC structural parameters of M3 are obtained from \cite{2018MNRAS.478.1520B}, where the online database is accessible at \url{https://people.smp.uq.edu.au/HolgerBaumgardt/globular/}. The \texttt{Corespray} Python package can be downloaded at \url{https://github.com/webbjj/corespray}.  The \texttt{dynesty} Python package \citep{2020MNRAS.493.3132S} is used to provide a functional form between escaped binary fraction, core density and escape velocity. This work also utilizes the \texttt{astropy} \citep{astropy:2022}, \texttt{numpy} \citep{harris2020array}, \texttt{scipy} \citep{2020SciPy-NMeth} and \texttt{matplotlib} \citep{Hunter:2007} Python packages.



\bibliographystyle{mnras}
\bibliography{example} 

\onecolumn
\begin{appendix}

\section{Logistic escaped binary fraction model parameter estimation} \label{sec:model}

\begin{figure*}
    \centering
    \includegraphics[width=0.85\textwidth]{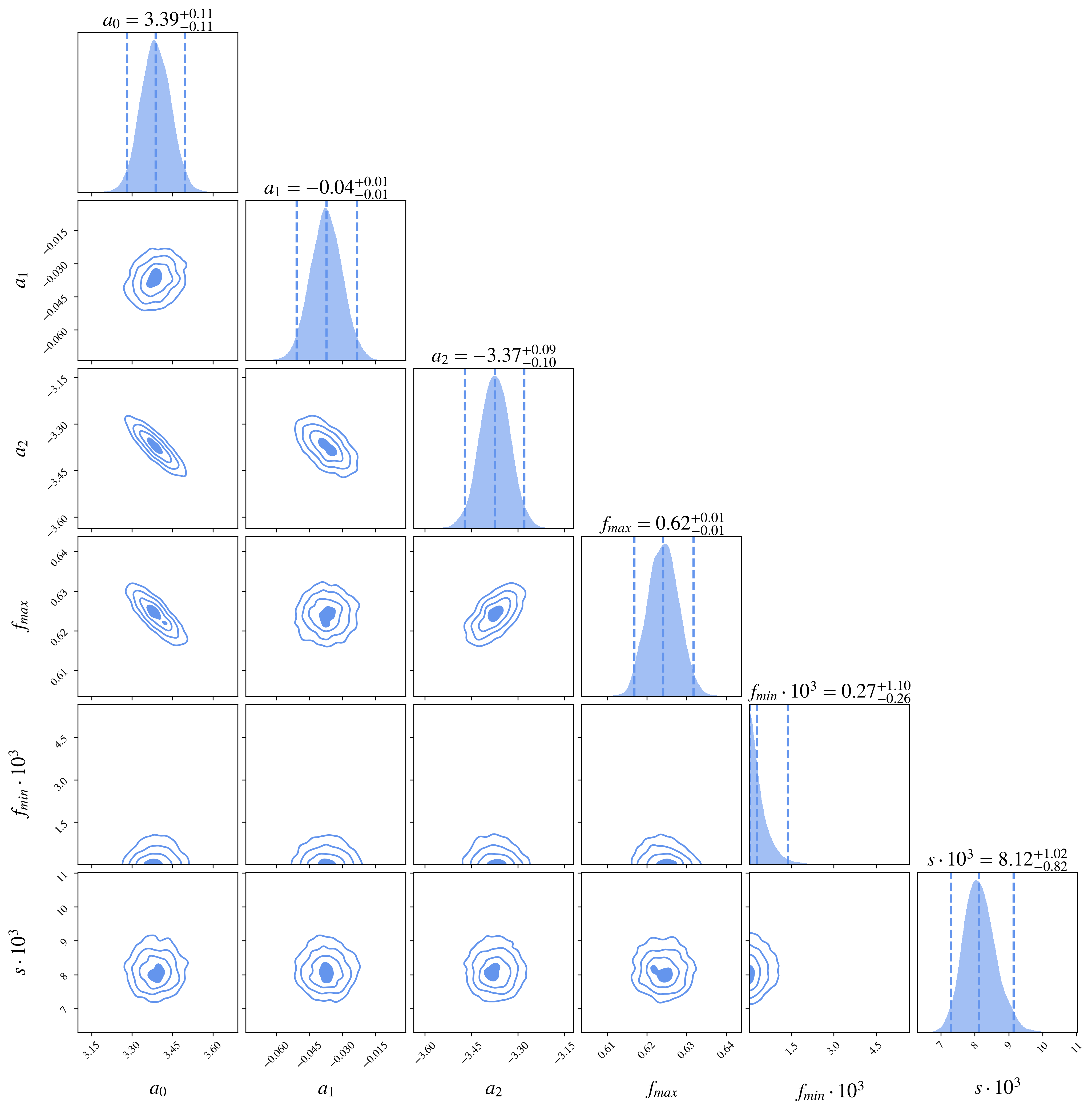}
    \caption{A cornerplot summarizing the best fit parameters for our logistic function modelling escaped binary fraction as a function of core density and escape velocity in Equation \ref{eq:funcform}. The free parameters $a_0$, $a_1$, and $a_2$ control the dependence of the escaped binary fraction on the core density and the escape velocity (the steepness and location of the curve), while $f_{\rm max}$ and $f_{\rm min}$ set the overall maximum and minimum allowed values. The best fit values  are derived by sampling from the posterior using \texttt{dynesty} \citep{2020MNRAS.493.3132S}. Note that $f_{min}$ and $s$ are multiplied by $10^{3}$ in this cornerplot only for visualization purposes. }
    \label{fig:cornerplot}
\end{figure*}

From Figure \ref{fig:binary-frac}, we observe a correlation between $\rho_c$, $v_{\rm esc}$, and escaped binary fraction ($f_{\rm b, esc}$). As discussed in Section \ref{sec:recoils}, we model $f_{\rm b, esc}$ as a function of $\rho_c$ and $v_{\rm esc}$ with a modified logistic function (Equation \ref{eq:funcform}). We fit for five free parameters in Equation \ref{eq:funcform}: $\theta=\{a_{0}, a_{1}, a_{2}, f_{max}, f_{min} \ \text{and} \ s\}$, where we originally assume broad uniform priors and determine the best fit parameters by sampling from the posterior using \texttt{dynesty} \citep{2020MNRAS.493.3132S} \texttt{v2.1.3}. Figure \ref{fig:cornerplot} shows a cornerplot that summarizes the posterior distributions (and uncertainties) for each of the free parameters.
       
\section{St\"{a}ckel approximation effects}
\label{sec:Staeckel}

As discussed in Sections \ref{sec:actions} and \ref{sec:potentialmodels}, actions in this study are computed using the St\"{a}ckel approximation in \texttt{galpy} \citep{2015ApJS..216...29B}. In a static potential, actions are a conserved quantity. However, in Figure \ref{fig:actionevolve} we see that the actions of individual GCs in the baseline \texttt{MWPotential2014} Galactic potential model have different actions at the beginning and end of the simulation. The clusters that exhibit the largest change in actions are those with small pericentre radii. The use of the St\"{a}ckel approximation in regions where the gradient in the Galactic potential is large, like the inner regions of the Galaxy, can lead to slightly different estimates of a cluster's orbit depending on its orbital phase and the degree to which its orbit precesses every orbital period. Hence, this shift is artificial and simply the result of our method for calculating actions.

To show this shift in actions as a function of pericentric radius, we compare the mean action coordinates of recent escapers to early escapers for our \texttt{Corespray} extra-tidal star distributions. For clusters with $ 5 \times P_{\text{orb}} < 12$Gyr, we compare stars that escaped within one $P_{\text{orb}}$ to those that escaped between four and five orbital periods in the past. For the 14 clusters listed in Section \ref{sec:gcsetup} with $ 5 \times P_{\text{orb}} > 12$Gyr, we compare the most recently escaped 10,000 \texttt{Corespray} extra-tidal stars to the mean action coordinates of the first 10,000 escapers. In Figure \ref{fig:St\"{a}ckel}, it is evident that there is a general trend of increasing shifts in action for decreasing pericentre radii. Clusters with pericentric radii smaller than five kiloparsecs typically exhibit the largest shifts in actions between recent and old escapers. Hence, these shifts are an artificial effect with inner clusters being more significant affected.

\begin{figure}
    \centering
    \includegraphics[width=0.7\columnwidth]{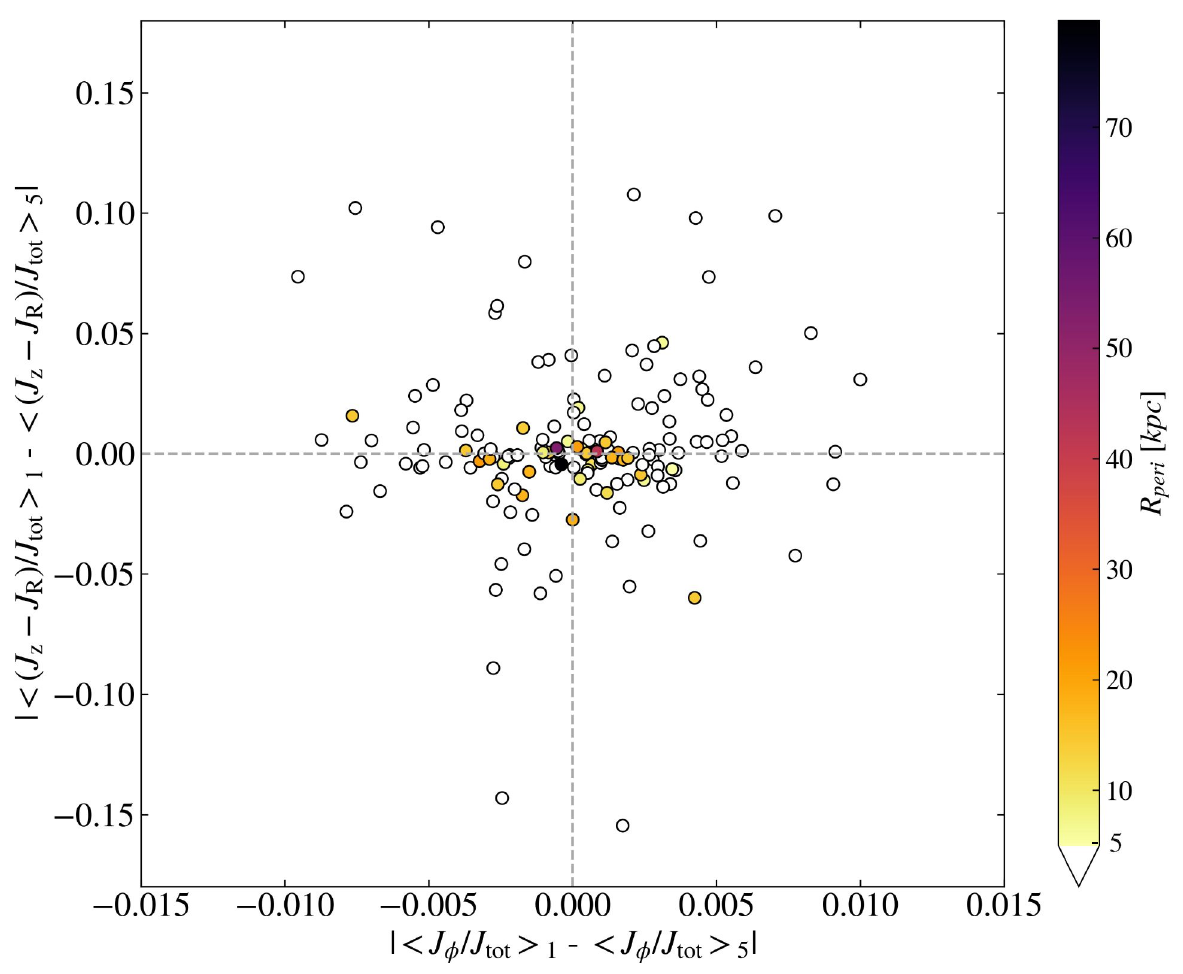}
    \caption{Action coordinate shifts between the means of the most recently and most early escaped \texttt{Corespray} extra-tidal stars in the baseline \texttt{MWPotential2014} Galactic potential model. For clusters with $ 5 \times P_{\text{orb}} < 12$Gyr, we compare stars that escaped within one $P_{\text{orb}}$ to those that escaped between four and five orbital periods in the past. For clusters with $ 5 \times P_{\text{orb}} > 12$Gyr, we compare the most recently escaped 10,000 \texttt{Corespray} extra-tidal stars to the mean action coordinates of the earliest 10,000 escapers. Each point is coloured by the host GC's pericentre radius, where clusters with pericentre radii less than five kiloparsecs are all plotted as white points. Zero-point shifts are plotted as dashed lines. Clusters with small pericentre radii typically exhibit larger action shifts among recent and older escapers.}
    \label{fig:St\"{a}ckel}
\end{figure}

\section{Action Coordinate Parameters}
\label{sec:params}
To roughly determine a host cluster association for a given extra-tidal star in the field, we list the action coordinates for each GC and the associated GEMS distribution in Table \ref{tab:actiontable}. Specifically, we present J$_{\phi}$ / J$_{\rm tot}$ and 
(J$_{\rm z}-$J$_{\rm R}$)/ J$_{\rm tot}$ for each GC analyzed in this study, with action coordinates computed using the orbital parameters in \cite{2018MNRAS.478.1520B} and the St\"{a}ckel approximation in \texttt{galpy} \citep{2015ApJS..216...29B}. The mean <J$_{\phi}$ / J$_{\rm tot}$> and <(J$_{\rm z}-$J$_{\rm R}$)/ J$_{\rm tot}$> values and corresponding standard deviations for the extra-tidal distributions are computed similarly (see Section \ref{sec:actions} for details). 



\begin{longtable}{ccccc}
\caption{Action coordinates for all GCs in \citet{2018MNRAS.478.1520B} and the corresponding GEMS distributions.} \label{tab:actiontable} \\
\multicolumn{1}{c}{\textbf{Globular Cluster}} & \multicolumn{1}{c}{J$_{\phi}$ / J$_{\rm tot}$ (GC)} & \multicolumn{1}{c}{(J$_{\rm z}-$J$_{\rm R}$)/ J$_{\rm tot}$ (GC)} & \multicolumn{1}{c}{<J$_{\phi}$ / J$_{\rm tot}$> $\pm \sigma$ (ET stars)} & \multicolumn{1}{c}{<(J$_{\rm z}-$J$_{\rm R}$)/ J$_{\rm tot}$> $\pm \sigma$ (ET stars)} \\ \hline \hline
\endfirsthead
\multicolumn{5}{c}
{{\bfseries \tablename\ \thetable{} -- continued from previous page}} \\
\multicolumn{1}{c}{\textbf{Globular Cluster}} & \multicolumn{1}{c}{J$_{\phi}$ / J$_{\rm tot}$ (GC)} & \multicolumn{1}{c}{(J$_{\rm z}-$J$_{\rm R}$)/ J$_{\rm tot}$ (GC)} & \multicolumn{1}{c}{<J$_{\phi}$ / J$_{\rm tot}$> $\pm \sigma$ (ET stars)} & \multicolumn{1}{c}{<(J$_{\rm z}-$J$_{\rm R}$)/ J$_{\rm tot}$> $\pm \sigma$ (ET stars)} \\ \hline \hline
\endhead
\hline \multicolumn{5}{r}{{Continued on next page.}} \\
\endfoot
\endlastfoot

2MASS-GC01 & 0.98 & -0.018 & 0.955 $\pm$ 0.067 & -0.034 $\pm$ 0.056\\
2MASS-GC02 & 0.289 & -0.504 & 0.269 $\pm$ 0.162 & -0.453 $\pm$ 0.164\\
AM\_1 & -0.103 & -0.703 & -0.093 $\pm$ 0.148 & -0.583 $\pm$ 0.255\\
AM\_4 & -0.28 & 0.611 & -0.273 $\pm$ 0.075 & 0.587 $\pm$ 0.128\\
Arp\_2 & 0.157 & 0.209 & 0.149 $\pm$ 0.124 & 0.182 $\pm$ 0.222\\
BH\_140 & 0.026 & -0.972 & 0.015 $\pm$ 0.229 & -0.707 $\pm$ 0.347\\
BH\_261 & 0.765 & 0.078 & 0.742 $\pm$ 0.096 & 0.057 $\pm$ 0.086\\
Crater & -0.141 & 0.658 & -0.135 $\pm$ 0.098 & 0.596 $\pm$ 0.232\\
Djor\_1 & 0.394 & -0.575 & 0.366 $\pm$ 0.193 & -0.539 $\pm$ 0.183\\
Djor\_2 & -0.751 & 0.04 & -0.714 $\pm$ 0.132 & 0.035 $\pm$ 0.111\\
ESO\_280-SC06 & 0.284 & -0.269 & 0.269 $\pm$ 0.125 & -0.263 $\pm$ 0.153\\
ESO\_452-SC11 & 0.036 & 0.4 & 0.034 $\pm$ 0.124 & 0.338 $\pm$ 0.184\\
E\_3 & 0.833 & 0.073 & 0.814 $\pm$ 0.074 & 0.064 $\pm$ 0.073\\
Eridanus & 0.067 & -0.097 & 0.063 $\pm$ 0.123 & -0.099 $\pm$ 0.234\\
FSR\_1716 & 0.828 & -0.003 & 0.802 $\pm$ 0.092 & -0.012 $\pm$ 0.084\\
FSR\_1735 & 0.613 & -0.163 & 0.579 $\pm$ 0.152 & -0.13 $\pm$ 0.146\\
FSR\_1758 & -0.568 & -0.227 & -0.529 $\pm$ 0.169 & -0.212 $\pm$ 0.181\\
HP\_1 & 0.076 & -0.913 & 0.065 $\pm$ 0.25 & -0.633 $\pm$ 0.324\\
IC\_1257 & -0.118 & -0.743 & -0.107 $\pm$ 0.179 & -0.66 $\pm$ 0.216\\
IC\_1276 & 0.899 & -0.072 & 0.869 $\pm$ 0.092 & -0.086 $\pm$ 0.081\\
IC\_4499 & -0.289 & 0.099 & -0.272 $\pm$ 0.137 & 0.084 $\pm$ 0.211\\
Laevens\_3 & 0.258 & -0.092 & 0.247 $\pm$ 0.111 & -0.087 $\pm$ 0.179\\
Liller\_1 & -0.411 & -0.552 & -0.246 $\pm$ 0.438 & -0.305 $\pm$ 0.301\\
Lynga\_7 & 0.786 & -0.065 & 0.758 $\pm$ 0.11 & -0.074 $\pm$ 0.096\\
NGC\_104 & 0.843 & 0.142 & 0.771 $\pm$ 0.123 & 0.097 $\pm$ 0.125\\
NGC\_1261 & -0.202 & -0.392 & -0.174 $\pm$ 0.191 & -0.305 $\pm$ 0.259\\
NGC\_1851 & -0.109 & -0.628 & -0.08 $\pm$ 0.279 & -0.373 $\pm$ 0.315\\
NGC\_1904 & -0.019 & -0.775 & -0.015 $\pm$ 0.242 & -0.553 $\pm$ 0.287\\
NGC\_2298 & -0.236 & -0.479 & -0.211 $\pm$ 0.189 & -0.41 $\pm$ 0.225\\
NGC\_2419 & 0.313 & -0.137 & 0.276 $\pm$ 0.19 & -0.114 $\pm$ 0.28\\
NGC\_2808 & 0.21 & -0.735 & 0.16 $\pm$ 0.326 & -0.503 $\pm$ 0.306\\
NGC\_288 & -0.331 & 0.05 & -0.31 $\pm$ 0.128 & 0.047 $\pm$ 0.174\\
NGC\_3201 & -0.603 & -0.26 & -0.554 $\pm$ 0.181 & -0.238 $\pm$ 0.188\\
NGC\_362 & -0.023 & -0.339 & -0.018 $\pm$ 0.189 & -0.174 $\pm$ 0.274\\
NGC\_4147 & -0.016 & -0.395 & -0.014 $\pm$ 0.129 & -0.288 $\pm$ 0.253\\
NGC\_4372 & 0.809 & -0.044 & 0.774 $\pm$ 0.11 & -0.059 $\pm$ 0.101\\
NGC\_4590 & 0.509 & -0.13 & 0.473 $\pm$ 0.159 & -0.122 $\pm$ 0.191\\
NGC\_4833 & 0.289 & -0.632 & 0.257 $\pm$ 0.221 & -0.568 $\pm$ 0.2\\
NGC\_5024 & 0.202 & 0.546 & 0.187 $\pm$ 0.14 & 0.476 $\pm$ 0.228\\
NGC\_5053 & 0.211 & 0.741 & 0.204 $\pm$ 0.094 & 0.697 $\pm$ 0.143\\
NGC\_5139 & -0.609 & -0.103 & -0.525 $\pm$ 0.227 & -0.119 $\pm$ 0.223\\
NGC\_5272 & 0.395 & 0.211 & 0.359 $\pm$ 0.157 & 0.186 $\pm$ 0.222\\
NGC\_5286 & -0.249 & -0.567 & -0.203 $\pm$ 0.273 & -0.423 $\pm$ 0.29\\
NGC\_5466 & -0.131 & -0.238 & -0.122 $\pm$ 0.139 & -0.213 $\pm$ 0.238\\
NGC\_5634 & 0.208 & -0.021 & 0.186 $\pm$ 0.161 & -0.005 $\pm$ 0.259\\
NGC\_5694 & -0.14 & -0.656 & -0.106 $\pm$ 0.257 & -0.338 $\pm$ 0.343\\
NGC\_5824 & 0.48 & 0.251 & 0.413 $\pm$ 0.188 & 0.178 $\pm$ 0.27\\
NGC\_5897 & 0.351 & 0.253 & 0.331 $\pm$ 0.117 & 0.235 $\pm$ 0.167\\
NGC\_5904 & 0.104 & -0.296 & 0.087 $\pm$ 0.198 & -0.211 $\pm$ 0.315\\
NGC\_5927 & 0.957 & 0.003 & 0.916 $\pm$ 0.086 & -0.019 $\pm$ 0.075\\
NGC\_5946 & 0.085 & -0.505 & 0.071 $\pm$ 0.22 & -0.421 $\pm$ 0.231\\
NGC\_5986 & 0.28 & -0.111 & 0.242 $\pm$ 0.201 & -0.125 $\pm$ 0.222\\
NGC\_6093 & 0.166 & 0.679 & 0.139 $\pm$ 0.143 & 0.495 $\pm$ 0.237\\
NGC\_6101 & -0.492 & -0.257 & -0.455 $\pm$ 0.168 & -0.229 $\pm$ 0.196\\
NGC\_6121 & 0.369 & -0.619 & 0.322 $\pm$ 0.239 & -0.496 $\pm$ 0.259\\
NGC\_6139 & 0.509 & 0.227 & 0.452 $\pm$ 0.174 & 0.181 $\pm$ 0.189\\
NGC\_6144 & -0.301 & 0.52 & -0.294 $\pm$ 0.1 & 0.484 $\pm$ 0.136\\
NGC\_6171 & 0.5 & 0.299 & 0.466 $\pm$ 0.122 & 0.25 $\pm$ 0.137\\
NGC\_6205 & -0.273 & 0.084 & -0.235 $\pm$ 0.179 & 0.077 $\pm$ 0.243\\
NGC\_6218 & 0.631 & 0.229 & 0.596 $\pm$ 0.11 & 0.199 $\pm$ 0.118\\
NGC\_6229 & 0.08 & -0.498 & 0.065 $\pm$ 0.205 & -0.327 $\pm$ 0.303\\
NGC\_6235 & 0.582 & 0.124 & 0.556 $\pm$ 0.12 & 0.109 $\pm$ 0.14\\
NGC\_6254 & 0.609 & 0.177 & 0.567 $\pm$ 0.132 & 0.145 $\pm$ 0.138\\
NGC\_6256 & 0.877 & 0.057 & 0.84 $\pm$ 0.102 & 0.039 $\pm$ 0.08\\
NGC\_6266 & 0.722 & 0.155 & 0.63 $\pm$ 0.193 & 0.047 $\pm$ 0.169\\
NGC\_6273 & -0.168 & 0.136 & -0.159 $\pm$ 0.186 & 0.134 $\pm$ 0.248\\
NGC\_6284 & -0.075 & -0.09 & -0.064 $\pm$ 0.166 & -0.006 $\pm$ 0.237\\
NGC\_6287 & 0.015 & 0.039 & 0.015 $\pm$ 0.187 & -0.119 $\pm$ 0.237\\
NGC\_6293 & -0.227 & 0.038 & -0.205 $\pm$ 0.176 & 0.066 $\pm$ 0.216\\
NGC\_6304 & 0.837 & 0.053 & 0.801 $\pm$ 0.108 & 0.037 $\pm$ 0.088\\
NGC\_6316 & 0.576 & -0.063 & 0.526 $\pm$ 0.182 & -0.117 $\pm$ 0.158\\
NGC\_6325 & -0.331 & 0.332 & -0.323 $\pm$ 0.142 & 0.276 $\pm$ 0.168\\
NGC\_6333 & 0.243 & -0.206 & 0.225 $\pm$ 0.213 & -0.21 $\pm$ 0.24\\
NGC\_6341 & 0.02 & -0.164 & 0.016 $\pm$ 0.183 & -0.065 $\pm$ 0.274\\
NGC\_6342 & 0.35 & 0.46 & 0.337 $\pm$ 0.116 & 0.437 $\pm$ 0.144\\
NGC\_6352 & 0.946 & 0.028 & 0.917 $\pm$ 0.071 & 0.014 $\pm$ 0.059\\
NGC\_6355 & -0.17 & 0.119 & -0.164 $\pm$ 0.167 & 0.061 $\pm$ 0.196\\
NGC\_6356 & 0.699 & 0.08 & 0.647 $\pm$ 0.14 & 0.056 $\pm$ 0.149\\
NGC\_6362 & 0.647 & 0.261 & 0.618 $\pm$ 0.096 & 0.234 $\pm$ 0.107\\
NGC\_6366 & 0.749 & -0.054 & 0.723 $\pm$ 0.1 & -0.067 $\pm$ 0.093\\
NGC\_6380 & -0.33 & -0.389 & -0.287 $\pm$ 0.22 & -0.229 $\pm$ 0.277\\
NGC\_6388 & -0.642 & -0.075 & -0.543 $\pm$ 0.232 & -0.087 $\pm$ 0.213\\
NGC\_6397 & 0.388 & -0.6 & 0.346 $\pm$ 0.232 & -0.491 $\pm$ 0.252\\
NGC\_6401 & -0.2 & -0.379 & -0.183 $\pm$ 0.199 & -0.27 $\pm$ 0.24\\
NGC\_6402 & 0.349 & 0.126 & 0.304 $\pm$ 0.191 & 0.056 $\pm$ 0.208\\
NGC\_6426 & 0.629 & -0.188 & 0.598 $\pm$ 0.137 & -0.183 $\pm$ 0.143\\
NGC\_6440 & -0.403 & 0.023 & -0.324 $\pm$ 0.256 & 0.016 $\pm$ 0.295\\
NGC\_6441 & 0.768 & -0.09 & 0.659 $\pm$ 0.215 & -0.123 $\pm$ 0.193\\
NGC\_6453 & 0.196 & 0.084 & 0.171 $\pm$ 0.162 & 0.189 $\pm$ 0.237\\
NGC\_6496 & 0.694 & 0.082 & 0.676 $\pm$ 0.099 & 0.075 $\pm$ 0.102\\
NGC\_6517 & 0.363 & -0.218 & 0.307 $\pm$ 0.242 & -0.208 $\pm$ 0.212\\
NGC\_6522 & 0.368 & 0.511 & 0.333 $\pm$ 0.14 & 0.435 $\pm$ 0.182\\
NGC\_6528 & 0.208 & -0.016 & 0.201 $\pm$ 0.202 & -0.108 $\pm$ 0.169\\
NGC\_6535 & -0.589 & -0.178 & -0.559 $\pm$ 0.14 & -0.179 $\pm$ 0.123\\
NGC\_6539 & 0.566 & 0.392 & 0.535 $\pm$ 0.109 & 0.357 $\pm$ 0.12\\
NGC\_6541 & 0.526 & 0.104 & 0.496 $\pm$ 0.153 & 0.123 $\pm$ 0.165\\
NGC\_6544 & 0.379 & -0.581 & 0.32 $\pm$ 0.262 & -0.44 $\pm$ 0.279\\
NGC\_6553 & 0.971 & -0.006 & 0.93 $\pm$ 0.085 & -0.03 $\pm$ 0.072\\
NGC\_6558 & 0.21 & 0.031 & 0.196 $\pm$ 0.191 & -0.078 $\pm$ 0.227\\
NGC\_6569 & 0.811 & 0.156 & 0.768 $\pm$ 0.105 & 0.133 $\pm$ 0.096\\
NGC\_6584 & 0.257 & -0.423 & 0.235 $\pm$ 0.18 & -0.352 $\pm$ 0.235\\
NGC\_6624 & 0.123 & 0.708 & 0.114 $\pm$ 0.103 & 0.597 $\pm$ 0.214\\
NGC\_6626 & 0.465 & -0.168 & 0.401 $\pm$ 0.229 & -0.116 $\pm$ 0.243\\
NGC\_6637 & 0.246 & 0.727 & 0.23 $\pm$ 0.096 & 0.658 $\pm$ 0.142\\
NGC\_6638 & 0.119 & -0.044 & 0.098 $\pm$ 0.212 & -0.136 $\pm$ 0.3\\
NGC\_6642 & -0.177 & -0.037 & -0.155 $\pm$ 0.228 & -0.178 $\pm$ 0.253\\
NGC\_6652 & 0.085 & 0.417 & 0.078 $\pm$ 0.127 & 0.374 $\pm$ 0.224\\
NGC\_6656 & 0.648 & -0.148 & 0.593 $\pm$ 0.179 & -0.151 $\pm$ 0.182\\
NGC\_6681 & -0.011 & -0.202 & -0.009 $\pm$ 0.173 & 0.012 $\pm$ 0.258\\
NGC\_6712 & -0.082 & -0.29 & -0.071 $\pm$ 0.191 & -0.288 $\pm$ 0.226\\
NGC\_6715 & 0.138 & 0.205 & 0.096 $\pm$ 0.271 & 0.056 $\pm$ 0.456\\
NGC\_6717 & 0.583 & 0.344 & 0.551 $\pm$ 0.11 & 0.301 $\pm$ 0.115\\
NGC\_6723 & 0.03 & 0.776 & 0.03 $\pm$ 0.091 & 0.69 $\pm$ 0.18\\
NGC\_6749 & 0.768 & -0.216 & 0.736 $\pm$ 0.135 & -0.229 $\pm$ 0.116\\
NGC\_6752 & 0.841 & 0.095 & 0.791 $\pm$ 0.11 & 0.067 $\pm$ 0.104\\
NGC\_6760 & 0.774 & -0.186 & 0.733 $\pm$ 0.142 & -0.2 $\pm$ 0.125\\
NGC\_6779 & -0.263 & -0.327 & -0.236 $\pm$ 0.191 & -0.272 $\pm$ 0.234\\
NGC\_6809 & 0.298 & 0.161 & 0.277 $\pm$ 0.139 & 0.144 $\pm$ 0.185\\
NGC\_6838 & 0.967 & -0.009 & 0.939 $\pm$ 0.073 & -0.024 $\pm$ 0.062\\
NGC\_6864 & 0.191 & -0.4 & 0.147 $\pm$ 0.255 & -0.251 $\pm$ 0.314\\
NGC\_6934 & 0.236 & -0.698 & 0.203 $\pm$ 0.237 & -0.527 $\pm$ 0.28\\
NGC\_6981 & 0.015 & -0.663 & 0.012 $\pm$ 0.178 & -0.521 $\pm$ 0.258\\
NGC\_7006 & -0.155 & -0.684 & -0.134 $\pm$ 0.21 & -0.512 $\pm$ 0.286\\
NGC\_7078 & 0.718 & 0.026 & 0.638 $\pm$ 0.174 & -0.006 $\pm$ 0.186\\
NGC\_7089 & -0.122 & -0.49 & -0.095 $\pm$ 0.252 & -0.302 $\pm$ 0.317\\
NGC\_7099 & -0.254 & 0.154 & -0.228 $\pm$ 0.152 & 0.135 $\pm$ 0.214\\
NGC\_7492 & -0.037 & -0.022 & -0.034 $\pm$ 0.119 & -0.034 $\pm$ 0.214\\
Pal\_10 & 0.907 & -0.08 & 0.865 $\pm$ 0.111 & -0.096 $\pm$ 0.098\\
Pal\_11 & 0.829 & 0.096 & 0.807 $\pm$ 0.073 & 0.084 $\pm$ 0.071\\
Pal\_12  & 0.234 & -0.022 & 0.223 $\pm$ 0.12 & -0.025 $\pm$ 0.193\\
Pal\_13 & -0.187 & -0.296 & -0.179 $\pm$ 0.109 & -0.268 $\pm$ 0.187\\
Pal\_14 & -0.016 & -0.895 & -0.014 $\pm$ 0.15 & -0.724 $\pm$ 0.267\\
Pal\_15 & 0.07 & -0.575 & 0.063 $\pm$ 0.122 & -0.493 $\pm$ 0.234\\
Pal\_1 & 0.95 & 0.025 & 0.928 $\pm$ 0.071 & 0.015 $\pm$ 0.061\\
Pal\_2 & 0.104 & -0.878 & 0.088 $\pm$ 0.243 & -0.663 $\pm$ 0.293\\
Pal\_3 & 0.236 & 0.625 & 0.223 $\pm$ 0.105 & 0.564 $\pm$ 0.227\\
Pal\_4 & -0.041 & -0.53 & -0.038 $\pm$ 0.131 & -0.466 $\pm$ 0.235\\
Pal\_5 & 0.381 & 0.549 & 0.372 $\pm$ 0.078 & 0.527 $\pm$ 0.109\\
Pal\_6 & -0.044 & -0.22 & -0.038 $\pm$ 0.212 & -0.182 $\pm$ 0.176\\
Pal\_8 & 0.675 & -0.042 & 0.643 $\pm$ 0.119 & -0.046 $\pm$ 0.11\\
Pyxis & -0.081 & -0.039 & -0.074 $\pm$ 0.12 & -0.049 $\pm$ 0.247\\
Rup\_106 & 0.377 & -0.339 & 0.355 $\pm$ 0.149 & -0.31 $\pm$ 0.183\\
Sagittarius\_II & 0.412 & 0.144 & 0.39 $\pm$ 0.121 & 0.119 $\pm$ 0.2\\
Ter\_10 & 0.285 & -0.079 & 0.25 $\pm$ 0.199 & -0.06 $\pm$ 0.235\\
Ter\_12 & 0.828 & -0.039 & 0.794 $\pm$ 0.112 & -0.052 $\pm$ 0.095\\
Ter\_1 & 0.734 & -0.256 & 0.675 $\pm$ 0.198 & -0.274 $\pm$ 0.155\\
Ter\_2 & -0.71 & 0.101 & -0.647 $\pm$ 0.177 & 0.076 $\pm$ 0.155\\
Ter\_3 & 0.729 & 0.205 & 0.711 $\pm$ 0.082 & 0.196 $\pm$ 0.079\\
Ter\_4 & 0.589 & 0.19 & 0.528 $\pm$ 0.187 & 0.13 $\pm$ 0.19\\
Ter\_5 & 0.513 & -0.381 & 0.426 $\pm$ 0.3 & -0.361 $\pm$ 0.219\\
Ter\_6 & -0.647 & -0.157 & -0.562 $\pm$ 0.247 & -0.17 $\pm$ 0.189\\
Ter\_7 & 0.105 & 0.108 & 0.098 $\pm$ 0.129 & 0.084 $\pm$ 0.231\\
Ter\_8 & 0.107 & 0.092 & 0.1 $\pm$ 0.131 & 0.07 $\pm$ 0.245\\
Ter\_9 & 0.497 & -0.408 & 0.44 $\pm$ 0.244 & -0.402 $\pm$ 0.19\\
Ton\_2 & 0.706 & 0.095 & 0.678 $\pm$ 0.108 & 0.084 $\pm$ 0.107\\
UKS\_1 & 0.18 & -0.791 & 0.161 $\pm$ 0.21 & -0.71 $\pm$ 0.186\\
VVV-CL001 & -0.559 & -0.378 & -0.519 $\pm$ 0.2 & -0.367 $\pm$ 0.165\\
Whiting\_1 & 0.173 & 0.33 & 0.167 $\pm$ 0.099 & 0.305 $\pm$ 0.184\\

\hline \hline
\caption{Action coordinates for all 159 GCs analyzed in this study and the corresponding GEMS distributions. The cluster J$_{\phi}$ / J$_{\rm tot}$ and 
(J$_{\rm z}-$J$_{\rm R}$)/ J$_{\rm tot}$ values are computed using the orbital parameters in \citet{2018MNRAS.478.1520B} and the St\"{a}ckel approximation in \texttt{galpy} \citep{2015ApJS..216...29B}. The mean and standard deviation of the extra-tidal star J$_{\phi}$ / J$_{\rm tot}$ and 
(J$_{\rm z}-$J$_{\rm R}$)/ J$_{\rm tot}$ values are computed using the GEMS parameters.}
\end{longtable}


\section{Associating Extra-Tidal Field Stars with Individual Globular Clusters}\label{sec:association}

Using $J_{R}$, $J_{\phi}$ and $J_{z}$ action coordinates from the GEMS catalogue, one can determine the probability that any given field star originated from a certain GC using the \texttt{Python} code presented below. Specifically, this code uses a Gaussian kernel density estimator to compute both (i) the most likely cluster that the extra-tidal star originated from and (ii) the logarithm of the odds ratio between the first and second most likely progenitor clusters. For consistency between data sets, it is important to compute action coordinates of field stars using the St\"{a}ckel approximation in \texttt{galpy} \citep{2015ApJS..216...29B}.

\newpage
\begin{lstlisting}[language=python]
import numpy as np
from scipy.stats import gaussian_kde
from scipy import stats
import pandas as pd
import os

#Set path and identify files in GEMS Catalogue
catpath='.'
catfiles=os.listdir('%s/ETstars/' % catpath)
filenames=np.array([])
clusters=np.array([])
for file in catfiles:
    if 'csv' in file:
        filenames=np.append(filenames,'./ETstars/' + file)
        clusters=np.append(clusters,file[:-12])
            
#Build a Gaussian KDE in action diamond Space for each globular cluster
kernels=[]

#Establish a 100x100 grid within the action diamond
xmin,xmax=-1.01,1.01
ymin,ymax=-1.01,1.01

X, Y = np.mgrid[xmin:xmax:100j, ymin:ymax:100j]
position = np.vstack([X.ravel(), Y.ravel()])

#Load in data for each globular cluster and generate Gaussian KDE
for file in filenames:
    
    data=np.loadtxt(file,delimiter=',',skiprows=1,dtype=str)
    jr,jp,jz=data[:,7].astype(float),data[:,8].astype(float),data[:,9].astype(float)

    #Calculate x-axis and y-axis values in the action diamond
    jtot=np.fabs(jr)+np.fabs(jp)+np.fabs(jz)
    jy=(jz-jr)/jtot
    jx=jp/jtot
    
    #Build Gaussian kernel
    kernel = stats.gaussian_kde(np.vstack([jx, jy]))
    kernels.append(kernel)
    
#Assume you have a star with jx, jy=(0.2, 0.2). Find the value of the probability distribution function of extra-tidal stars for each cluster
jx_star,jy_star=0.2,-0.2

pdf=np.array([])
for k in kernels:
    pdf=np.append(pdf,k.pdf((jx_star,jy_star)))
    
#Find cluster that the star is most likely to have originated from given its actions
pdf_sort=np.argsort(pdf)
cluster_match=clusters[pdf_sort[-1]]
pdf_match=pdf[pdf_sort[-1]]
lpdf_ratio=np.log(pdf[pdf_sort[-1]]/pdf[pdf_sort[-2]])

print('This extra tidal star most likely came from %s, with the PDF reading %f at that location.' % (cluster_match,pdf_match))
print('The logarithm of the odds ratio between the first and second most likely progenitor clusters is %f' % lpdf_ratio)
\end{lstlisting}

\end{appendix}
\label{appendix:appendix}

\bsp	
\label{lastpage}
\end{document}